\documentclass[fleqn,usenatbib]{mnras}

\usepackage{newtxtext,newtxmath}
\usepackage[T1]{fontenc}
\usepackage{ae,aecompl}

\usepackage{graphicx}	% Including figure files
\usepackage{amsmath}	% Advanced maths commands
\usepackage{amssymb}	% Extra maths symbols
\usepackage{float}	% figure and stuff
\usepackage{multirow}

\title[ELT MCAO many-core CPU RTC]{A many-core CPU prototype of an MCAO and LTAO RTC for ELT-scale instruments}

\author[D. R. Jenkins et al.]{
David R. Jenkins,\thanks{E-mail: d.r.jenkins@durham.ac.uk}
Alastair G. Basden,
and Richard M. Myers
\\
CfAI, Department of Physics, Durham University, DH1 3LE, UK
}

\date{Accepted XXX. Received YYY; in original form ZZZ}

\pubyear{2018}

\begin{document}
\label{firstpage}
\pagerange{\pageref{firstpage}--\pageref{lastpage}}
\maketitle

\begin{abstract}
We propose a many-core CPU architecture for Extremely Large Telescope (ELT) scale adaptive optics (AO) real-time control (RTC) for the multi-conjugate AO (MCAO) and laser-tomographic AO (LTAO) modes. MCAO and LTAO differ from the more conventional single-conjugate (SCAO) mode by requiring more wavefront sensor (WFS) measurements and more deformable mirrors to achieve a wider field of correction, further increasing the computational requirements of ELT-scale AO. We demonstrate results of our CPU based AO RTC operating firstly in SCAO mode, using either Shack-Hartmann or Pyramid style WFS processing, and then in MCAO mode and in LTAO mode using the specifications of the proposed ELT instruments, MAORY and HARMONI. All results are gathered using a CPU based camera simulator utilising UDP packets to better demonstrate the pixel streaming and pipe-lining of the RTC software. We demonstrate the effects of switching parameters, streaming telemetry and implicit pseudo open-loop control (POLC) computation on the MCAO and LTAO modes. We achieve results of $<600\mu s$ latency with an ELT scale SCAO setup using Shack-Hartman processing and $<800\mu s$ latency with SCAO Pyramid WFS processing. We show that our MCAO and LTAO many core CPU architecture can achieve full system latencies of $<1000\mu s$ with jitters $<40\mu s$ RMS. We find that a CPU based AO RTC architecture has a good combination of performance, flexibility and maintainability for ELT-scale AO systems.

% It should be a single paragraph not more than 250 words (200 words for Letters).
\end{abstract}


%%%%%%%%%%%%%%%%%%%%%%%%%%%%%%%%%%%%%%%%%%%%%%%%%%

%%%%%%%%%%%%%%%%% BODY OF PAPER %%%%%%%%%%%%%%%%%%

% !TEX root = ../main.tex
\section{Introduction}

Ground based astronomical telescopes, both current and future, are much larger than space telescopes and so provide greater light collecting area and resolving power, increasing the potential for astronomical imaging. However due to the Earth's atmosphere, imaging from the ground without any active correction results in a significant degradation of diffraction limited imaging quality. Adaptive optics \citep[AO,][]{Babcock1953} is a widely-used technique that helps to negate the perturbing effects of the atmosphere and allows telescopes to achieve imaging fidelity much closer to the diffraction limit than otherwise. AO has been successfully used on sky and can be used in different configurations depending on the type and amount of correction needed.

For the next generation of extremely large telescopes (ELTs) such as the Giant Magellan Telescope \citep[GMT,][]{Johns2004}, the Thirty Meter Telescope \citep[TMT,][]{Stepp2004} and the Extremely Large Telescope \citep[ELT,][]{Spyromilio2008}, AO correction will become much more important but also much more difficult to achieve. The atmosphere affects imaging by reducing the effective diffraction limit of ground-based telescopes by perturbing the incoming wavefront and therefore ``blurring'' the image point-spread function (PSF). The relation between the strength of the atmospheric turbulence, and therefore the amount of wavefront perturbation, to the size of the PSF is known as the ``seeing'' limit.

The strength of the turbulence in the atmosphere which causes the wavefront perturbations can be defined by the Fried parameter \citep{Fried1966}, $r_0$, which has units of length and is usually on the order of \textasciitilde $10cm$ at a wavelength of around $500nm$. The Fried parameter is related to the size of the PSF in the seeing limit in a very similar way to the relationship of the telescope diameter to the size of the PSF in the diffraction limit. Therefore the effective diffraction limit for all telescopes with aperture diameters greater than $r_0$ will be constant for certain seeing conditions and roughly equal to the diffraction limit of a telescope with diameter $r_0$. This means that without any AO correction the resolving power of larger and larger ground-based telescopes remains effectively constant at the particular seeing limit.

\subsubsection{AO functionality}

The functionality of current AO systems can be split into 3 main parts: indirectly detecting the incoming wavefront, reconstructing the wavefront, and then applying corrections to mitigate the effects of the atmosphere. The detection and correction of the wavefront are usually performed by optical methods, using a wavefront sensor (WFS) for the detection and a deformable mirror (DM) for correction, and the reconstruction is a computational method. The basic idea behind the reconstruction is to attempt to ``flatten'' the wavefront from a point-source natural guide star (NGS) which picks up aberrations as it travels through the atmosphere. Therefore any deviations of the measured phase of the wavefront from a flat wavefront are considered perturbations and can be corrected by the DM.

For the reconstruction, analysis of the WFS images provides local wavefront gradients which are used to reconstruct the shape of the wavefront phase as projected onto the DM, this information is then used to produce commands for the DM to correct for the atmospheric perturbations. As the atmosphere is constantly changing, the corrections need to be applied very shortly after measurement for them to still be valid. This requires corrections to be performed in real-time; i.e. there is a defined time limit between measurement and correction within which the reconstruction must be computed and applied such that the AO performance is sufficient for the atmospheric conditions. The real-time control (RTC) of AO is therefore fundamental to the operation of AO and is an extremely important consideration in the design of an AO instrument.

The most basic form of AO is single conjugate AO (SCAO) using a single WFS on a natural guide star (NGS) to provide wavefront measurements for a single DM. An NGS is simply a bright star that is close enough to the science object of interest such that the light from both travels through as much of the same atmosphere as possible to ensure that the reconstruction along the NGS line of sight is also valid for the science object. The corrected field of view for SCAO is small and therefore it is greatly limited by sky-coverage; NGS need to be bright and close enough to the science target to reduce the effects of anisoplanatism \citep{Fried1982}. To overcome the problem of anisoplanatism, several more complex types of AO have been proposed to allow greater sky coverage and also to increase the corrected field of view.

\subsection{Multi-conjugate and laser-tomographic AO for ELTs}
\label{sec:mcao_ltao}

\begin{figure}
\includegraphics[width=1\columnwidth]{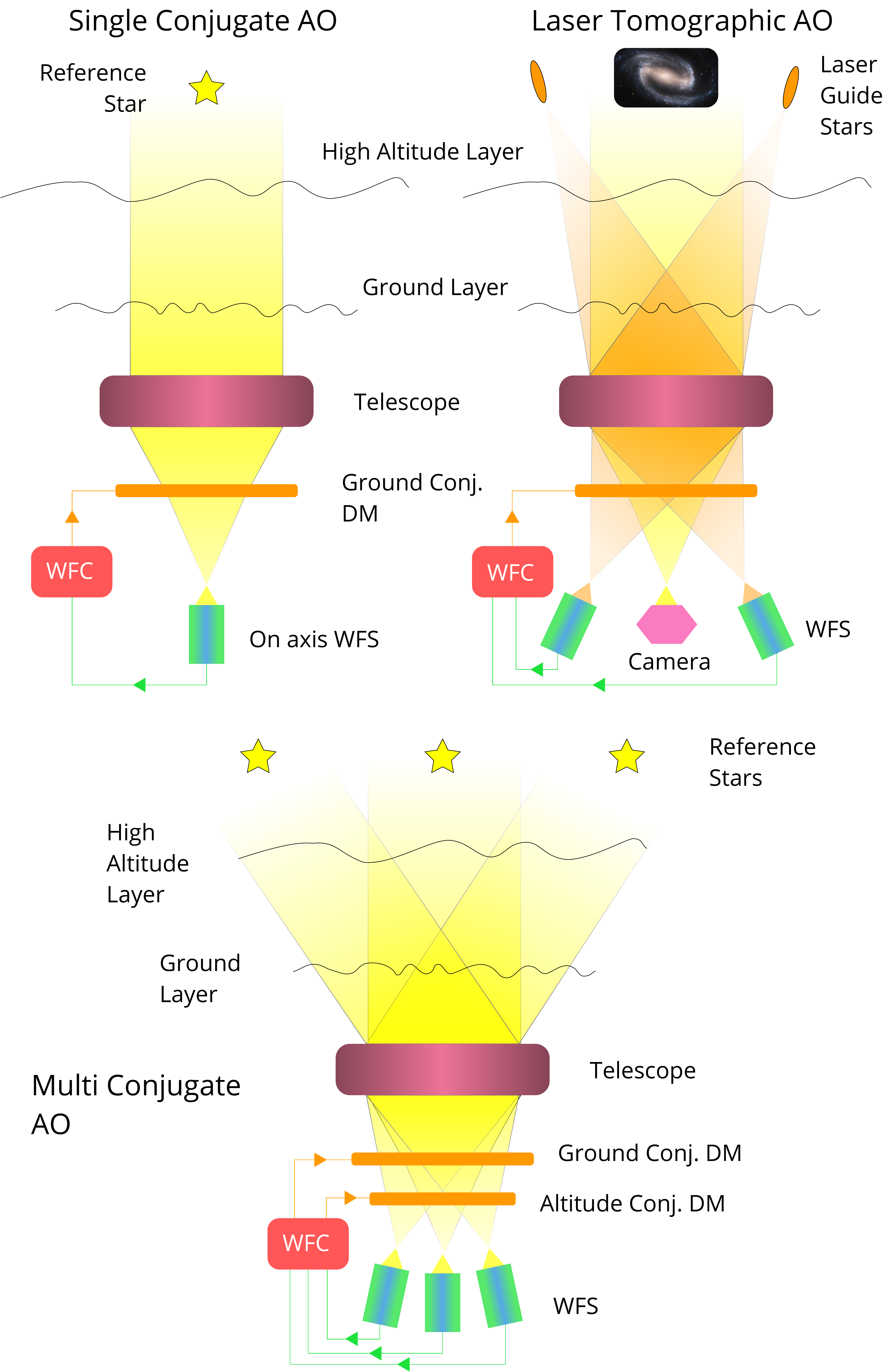}
\caption{A visual comparison of SCAO, LTAO and MCAO. SCAO in general has one WFS and one DM. MCAO has multiple WFSs focussed on different guide stars and multiple DMs conjugated to different atmospheric layers. LTAO is similar to MCAO except it mainly uses LGS, it generally only has one DM and it corrects over a narrower FoV. }
\label{fig:mcao}
\end{figure}

\begin{figure*}
    \includegraphics[width=1.8\columnwidth]{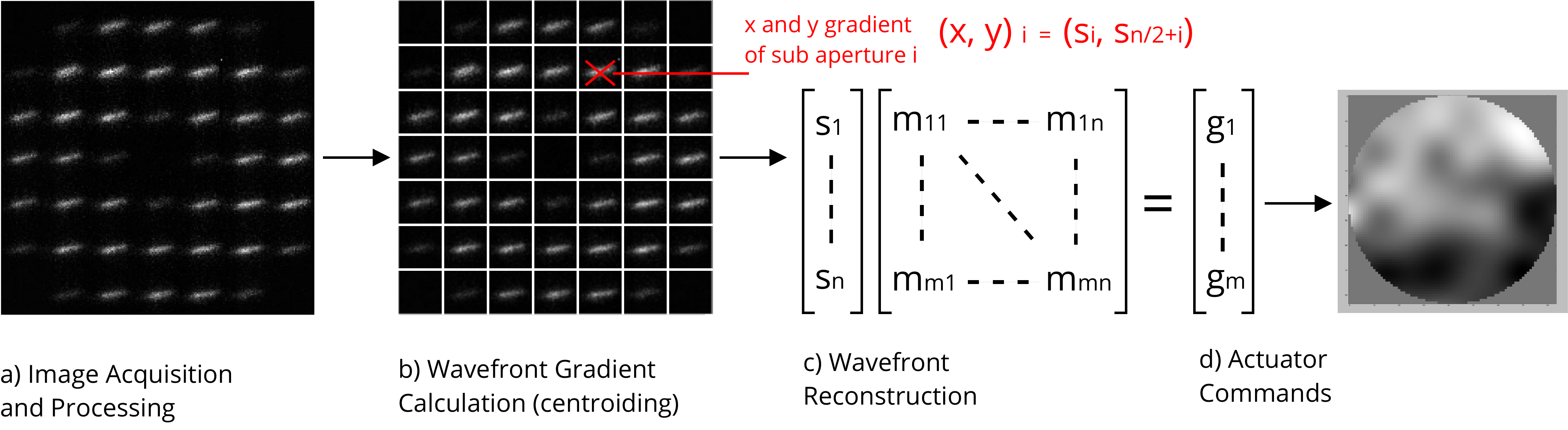}
    \caption{The four main processes in the AO loop for a single WFS. For the MCAO or LTAO case the final actuator commands from each WFS then need to be combined in a final step. }
    \label{fig:ao_diag}
    \end{figure*}

Multi conjugate AO \citep[MCAO,][]{Beckers1988} and laser tomographic \citep[LTAO,][]{,,Foy1985,Fugate1991} differ from the single conjugate AO (SCAO), described above, by having multiple DMs and/or multiple WFSs which increase the AO functionality over SCAO by having a wider field of view, increasing PSF field stability and the ability to do science on multiple objects. Figure \ref{fig:mcao} summarises the differences between SCAO, MCAO and LTAO.

MCAO gets its name from having multiple DMs conjugate to different heights in the atmosphere with multiple WFSs looking at different guide stars which can be either natural guide stars (NGS) or artificial laser guide stars (LGS). LGS use lasers to project a bright source of light high up in the atmosphere. This gives greater sky coverage as there is no fundamental restriction on the laser launch direction however it introduces a new source of error known as the cone effect or focus anisoplanatism \citep{Beckers1988}. LTAO relies mostly on LGS WFS and usually has a single DM conjugate to the ground layer of turbulence.

There is a another type of wide-field AO known as Multi-object AO \cite[MOAO,][]{MOAO2007} which also uses multiple WFSs and DMs, MOAO aims to correct the atmospheric perturbations along a line of sight for each science target. Therefore it uses one DM per science target in an open loop configuration with multiple guide stars across the total field. Due to the increased complexity of MOAO and its open loop nature, the architecture for MCAO/LTAO proposed here will not be directly applicable to MOAO without modification and so it will not be considered here.

The computational complexity of the reconstruction depends on both the spatial frequency of the wavefront-gradient sampling of the WFS and the number of degrees of freedom of the DM. These values both typically scale to the square of telescope diameter and therefore the computational complexity of the reconstruction scales to the fourth power of telescope diameter; becoming much more of a challenge for ELT-scale telescopes and especially for the more complex types of AO which can utilise multiple WFSs and DMs and also employ more demanding reconstruction algorithms. Each extra WFS and DM in MCAO and LTAO increases the number of parameters and degrees of freedom (DoF) such that the computational complexity scales linearly with the number of each compared with an SCAO system.

Previous investigations into using many-core CPU systems, specifically the Xeon Phi, for accelerating SCAO-like AO systems can be found in \citet{Barr2015} and \citet{Jenkins2018}. This report will continue the work in \citet{Jenkins2018} and apply the method to develop a prototype system suitable for the more complex and computationally demanding MCAO and LTAO. Section 2 will give a brief update on the general optimisations and SCAO results presented \citet{Jenkins2018}. Section 3 will describe the extension of this work to prototype an MCAO/LTAO system. Section 4 will present the results of this work and section 5 will conclude with a discussion of the results and future work.
% !TEX root = ../main.tex

\section{Recent Developments of DARC for many-core CPUs} \label{sec:devel_RTC}

The general optimisations and modifications made to the RTC software used in this report, the Durham Adaptive Optics Real-time Controller (DARC), is detailed in \citet{Jenkins2018}. However since then improvements have been made to the software resulting in an improvement in performance and functionality. One of the main changes is a move away from the Aravis GigE Vision Library \citep{Aravis} for simulating the camera and for receiving the pixel stream. We have instead implemented a camera simulator based on the proposed standard for ESO ELT WFS \citep{Downing2018} that streams pixels using UDP packets, and we have also developed a library within the RTC to receive this new camera stream.

\subsection{UDP Camera Simulator}

We have developed the new ESO camera interface from the ground up, rather than relying on a pre-existing library which is not optimised for the extreme performance we require. The UDP camera simulator is controlled at runtime of the simulator executable and the receiver software simply waits to receive the packets. In this way we can define the parameters of the camera simulator separately from the receiving of the camera stream, and it no longer requires a heartbeat thread to keep the camera sending packets, which we found could interfere with the pixel stream.

The camera simulator is implemented in the C programming language in an effort to make it both as low level as possible and as easy to modify and develop as possible. The underlying networking uses packet sockets, which are used to receive or send raw packets at the device driver (OSI Layer 2) level \citep{PacketSocket}. This in theory allows minimal overhead from the kernel when sending and receiving packets as most of the protocol implementation can be programmed in user space on top of the physical layer. The type of socket used here is SOCK\_DGRAM which does not have the link level header removed by the network stack and so is not quite as low level as SOCK\_RAW packets.

The operating system on the simulator machine is Ubuntu 16.04 on top of a Linux 4.4.0 generic kernel. A low-latency Linux kernel was investigated but was observed to provide no discernible performance benefit. Some OS, Kernel and network level tuning was performed to improve performance. Some of the steps taken involved:
\setlength{\leftmargini}{ .18in}
\renewcommand{\labelitemi}{$\circ$}
\begin{itemize}
    \setlength{\itemindent}{-.009in}
    \item isolating most CPU cores from the OS scheduler,
    \item specifying the core affinity for the NIC interrupts and camera threads,
    \item tuning both the NICs and linux network stacks UDP buffer sizes using the linux ``ethtool'' command and the ``sysctl'' utility,
    \item and setting the CPU power settings to ``performance''.
\end{itemize}

The camera simulator software is used only to simulate the pipe-lined transfer of pixels to the RTC and so the unique images that were streamed by each camera were created ahead of time via AO simulations to properly construct images for the type and dimensions of AO system tested. The images were stored on a PCIe fast raid storage array consisting of 4 Samsung 960 EVO NVMe solid state drives (SSDs) which provided transfer rates $>4.2GBs^{-1}$ which is sufficient to allow the pixel data to be streamed directly from storage. The simulator software is set up such that both the inter-packet delay and inter-frame delay can be set independently. This not only allows different camera frame rates to be simulated, it also allows the readout time to be adjusted to better reflect that of a real camera, rather than just sending out the packets as soon as possible. As the inter-packet delay needs to have microsecond precision the timing is achieved via the Linux ``timer\_fd'' utility which uses file descriptors to achieve a repeatable high precision timer.

The simulator hardware consists of a 2012 Intel Xeon E5-2650 dual socket system with 8 CPU cores and 32GB of DDR3-1600Mhz RAM per socket with a base CPU frequency of 2.0GHz. The network devices used are 2 PCIe Intel Ethernet X710-DA4 Network Interface Controllers (NICs) with 4 10GbE ports each where a single camera simulator stream will have exclusive use of one of these 8 interfaces. In a multi-socket CPU system, certain PCI-e lanes are physically connected to a single CPU socket. The NICs were therefore installed in the host such that each was local to a different CPU socket and then the camera threads for each interface were assigned CPU cores on the local socket.

% Some BIOS setting were tuned for the simulator, Intel HyperThreading was turned off to allow a single processing thread exclusive use of a CPU hardware core and the power setting were tuned to performance settings. Linux kernel settings were tuned such that the relevant command line options were ``isolcpus=2-15 nohz\_full=2-15 nohalt idle=poll''. The ``isolcpus'' option isolates all but the first 2 CPU cores from the OS scheduler such that processes must be explicitly allocated to them, this prevents the OS from potentially interrupting the simulator processes. The ``nohz\_full'' option sets the specified CPUs whose tick will be stopped whenever possible, this can reduce the number of scheduling-clock interrupts and reduce jitter. The ``nohalt'' option tells the kernel not to use certain power saving functions which reduces interrupt wake-up latency and can improve performance for real-time systems. Finally the ``idle=poll'' option forces a polling idle loop that can slightly improve the performance of waking an idle CPU at the expense of power. Descriptions of kernel command line parameters can be found at \cite{kernelcommandline}.

\begin{figure}
    \includegraphics[width=1\columnwidth]{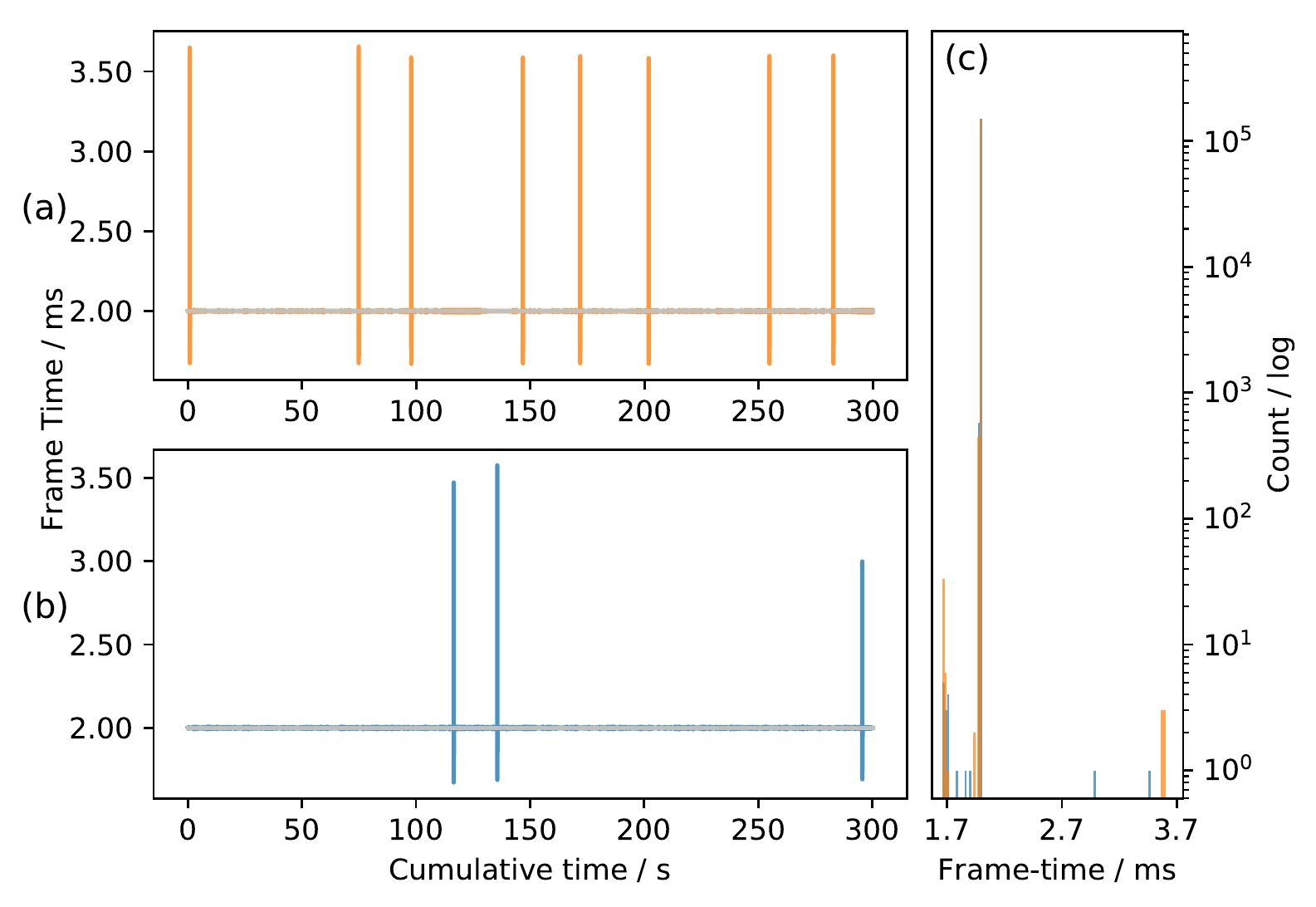}
    \caption{Frametime results from the camera simulator software showing two representative samples of frametime distributions while the camera simulator is delivering the 7 individual camera streams as needed by the MCAO and LTAO RTC architectures described in Section~\ref{sec:mcao_ltao}. Results shown are for $1.5\times10^5$ iterations at $500Hz$ for a total time of $300s$.}
    \label{fig:camera_lat}
\end{figure}

The camera simulator is unable to provide completely jitter free cameras streams, Figure~\ref{fig:camera_lat} shows two representative frame time distributions from the camera simulator whilst it was delivering 7 individual camera streams for an MCAO or LTAO setup as described in Section~\ref{sec:mcao_ltao}. For 10 random distributions similar to those shown, the number of outliers can vary from 0-11 over the 300s of running time, the largest outlier is never more than twice the frame time. For the periods of low jitter the RMS jitter is very low and on order of $5\mu s$. Each of these data sets were collected after restarting the camera simulator software and the amount of jitter present in each run can vary significantly. This introduces random high latency spikes into some of the AO RTC timing data presented in this report. This was minimised by waiting a small amount of time to determine if a certain run was more stable than normal. Once the software was running the amount of jitter varied little and so once a stable run was found, it was used for as many RTC tests as possible.

Due to the simulator software needing to time the inter-packet delay to microsecond precision and to deliver up to 7 individual camera streams at high frame-rate, the CPU system used is not an ideal candidate. The workload of the simulator is very different to that of the AO RTC system and so a more modern single socket CPU system with faster cores and lower latency memory could potentially improve the camera simulator's jitter performance. We note that the jitter originating from the camera simulator would not be present in real cameras as they are usually deterministic.

\subsection{Pyramid WFS, Pixel Handling and Slope Computation}

We have also implemented a SCAO RTC demonstrator for a system using a Pyramid WFS (Pyr-WFS). Pyr-WFSs \citep{Ragazzoni1996} differ from the more conventional Shack-Hartman WFS (SH-WFS) in the fundamental approach to detecting the local tip/tilt slopes across the telescope aperture. In the context of the RTC the main difference between the slope computation of a Pyr-WFS to a SH-WFS is the reduced pixel counts in the calculation of each slope measurement; only 4 pixels are needed for each slope measurement with a Pyr-WFS, whereas the SH-WFS generally requires more ($>36$ for ELT-scale) depending on the specific AO systems requirements.

\begin{figure}
    \includegraphics[width=1\columnwidth]{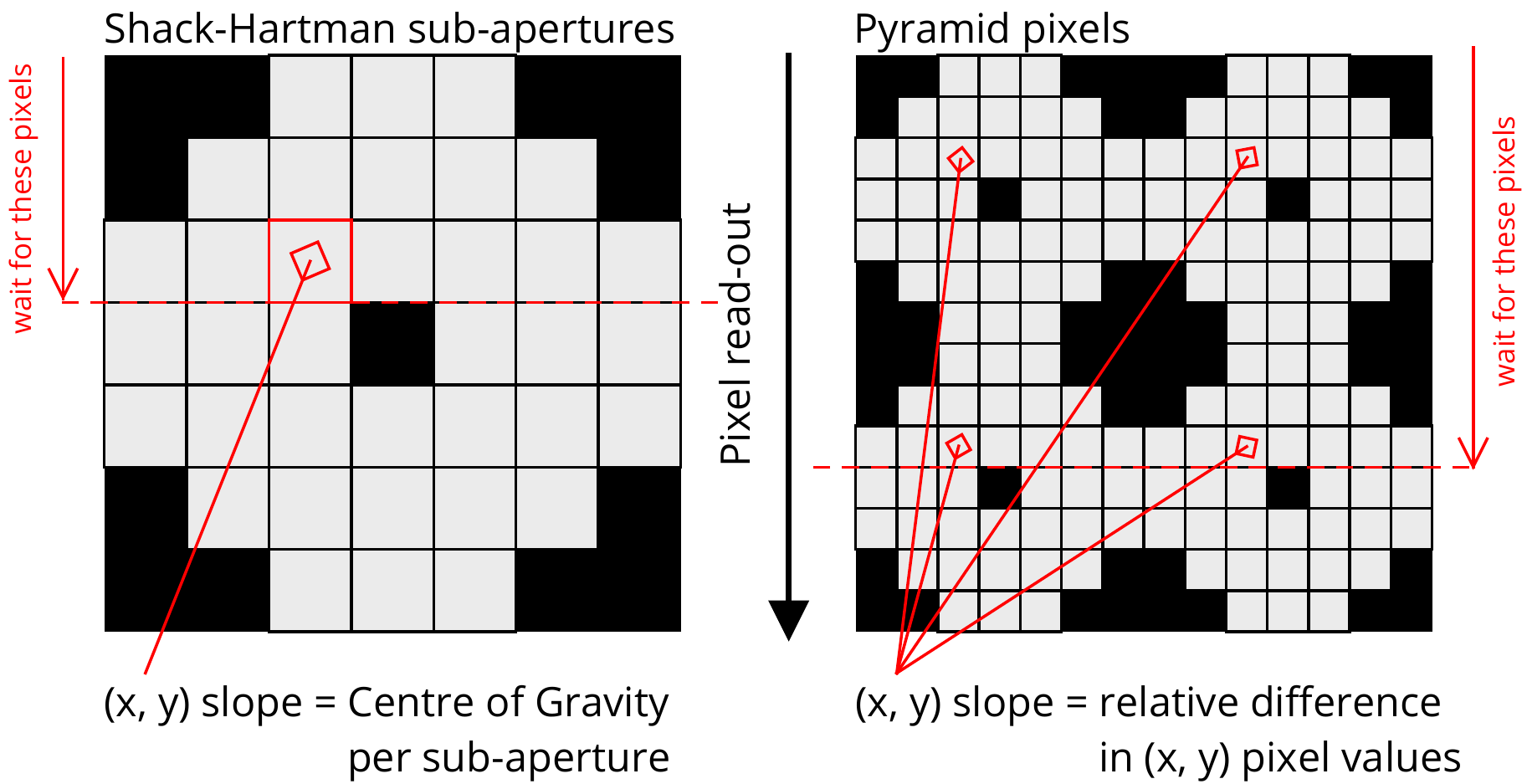}
    \caption{Comparison of the pixel pipe-lining for SH-WFS and Pyramid WFS image layouts. Each slope measurement from a SH-WFS is taken as the centre of gravity of the pixels values in each sub-aperture, slopes can be calculated as soon as all sub-aperture pixels have arrived. For the Pyramid WFS the slopes are calculated as the relative x and y differences between corresponding pixels from each pupil quadrant, no slopes can be computed until at least half the frame has been read. This difference makes pipe-ling of slopes much less effective when using a Pyramid WFS. Here we have assumed that the pixels are read from the top of the detector to the bottom, however the principle is the same for other read-out regimes. }
    \label{fig:wfs_slopes}
\end{figure}

There is also a difference in the memory access of the pixels required for each slope measurement and in the pipe-lining of the pixels, shown in Figure~\ref{fig:wfs_slopes}. For a Pyr-WFS the RTC needs to wait for more than half the WFS image to arrive before it can begin slope calculation whereas for a SH-WFS centroiding can begin as the soon as enough lines of pixels have arrived to complete a row of sub-apertures. This shows that even though there are far fewer pixels to process for a Pyr-WFS, the less efficient pipe-lining means that processing Pyr-WFSs in the RTC can be less computationally efficient than SH-WFSs.

\subsection{Up to date SCAO results with camera simulator}\label{sec:scaoresults}

\begin{figure}
    \includegraphics[width=1\columnwidth]{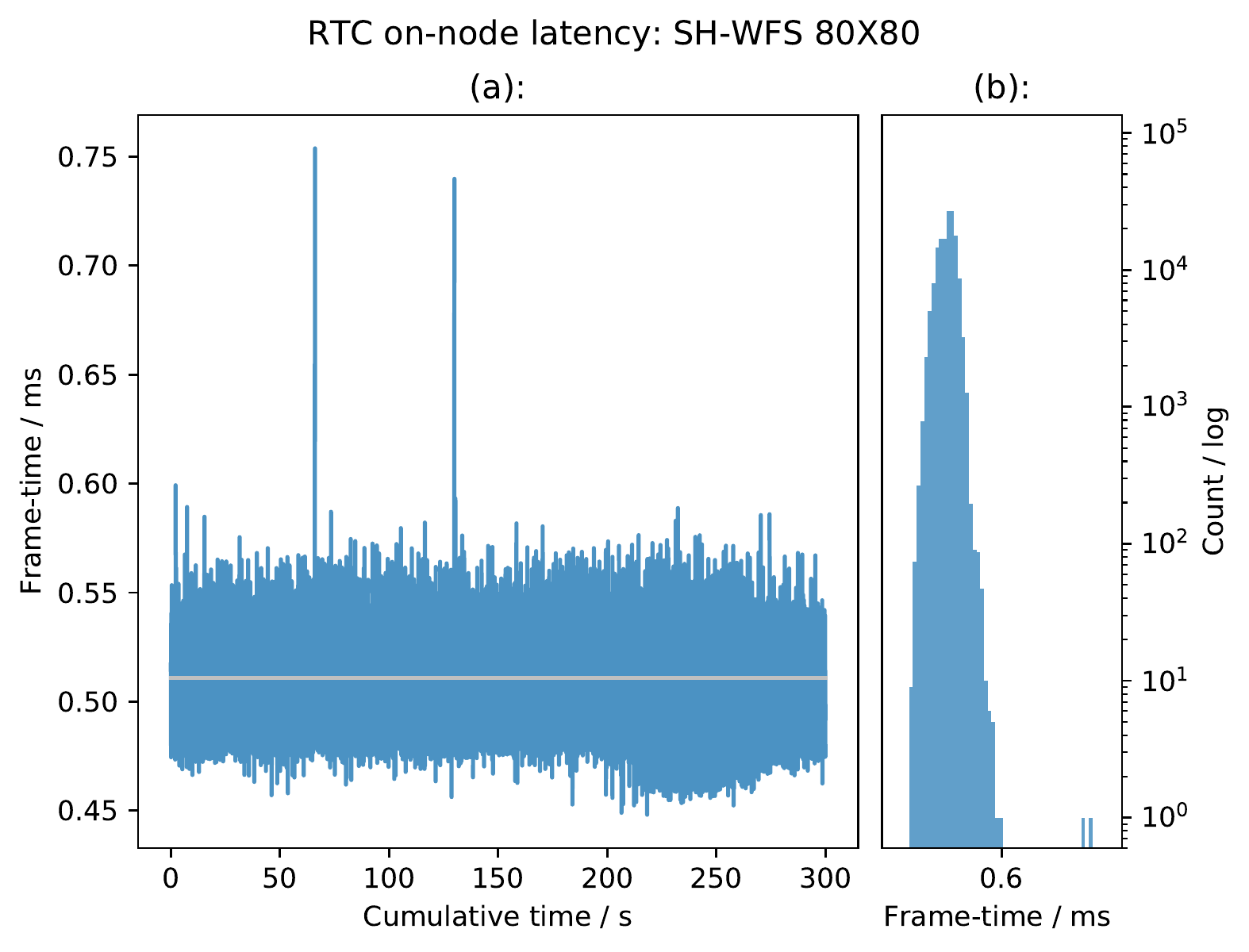}
    \caption{Frametime results for an SCAO setup using a Shack-hartman type WFS slope calculation with $80$ subapertures across the pupil. Results shown are for $1.5\times10^5$ iterations at $500Hz$ for a total time of $300s$. The mean latency is $511 \pm 15 \mu s$. }
    \label{fig:shwfs_scao}
\end{figure}

\begin{figure}
    \includegraphics[width=1\columnwidth]{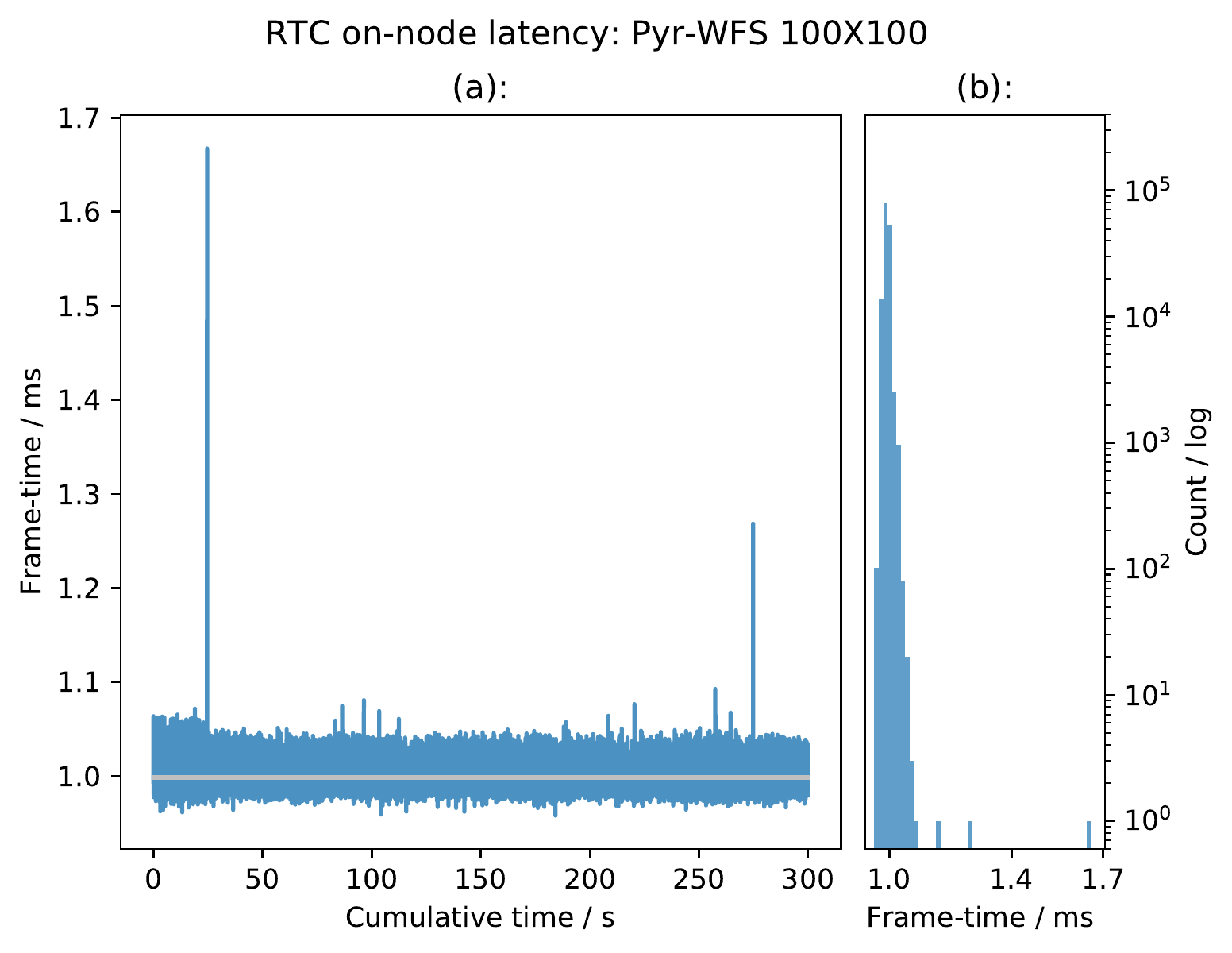}
    \caption{Frametime results for an SCAO setup using a Pyramid type WFS slope calculation with $100$ pixels across each quadrant. Results shown are for $1.5\times10^5$ iterations at $500Hz$ for a total time of $300s$. The mean latency is $998 \pm 10 \mu s$. }
    \label{fig:pyrwfs_scao}
\end{figure}

\begin{figure}
    \includegraphics[width=1\columnwidth]{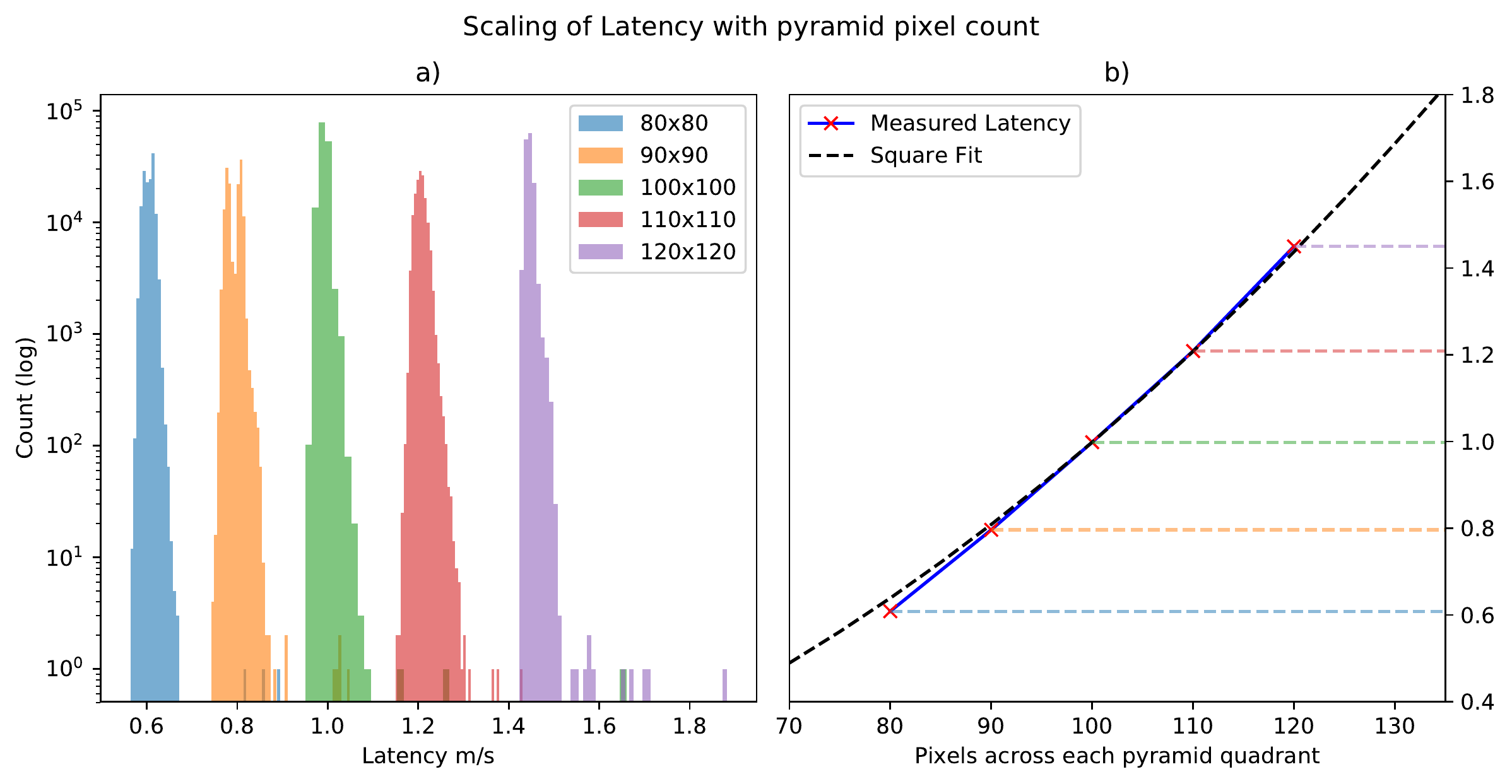}
    \caption{The scaling of latency with pixels across the pupil for the Pyramid type WFS. The latency is that for the entire RTC operation from pixels to DM commands as shown in Figure~\ref{fig:ao_diag}, measured from the time the last pixel arrives until the DM command is ready. The relevant values for the data are shown in Table~\ref{tab:ao_results}. }
    \label{fig:pyr_scaling}
\end{figure}

\begin{figure}
    \includegraphics[width=1\columnwidth]{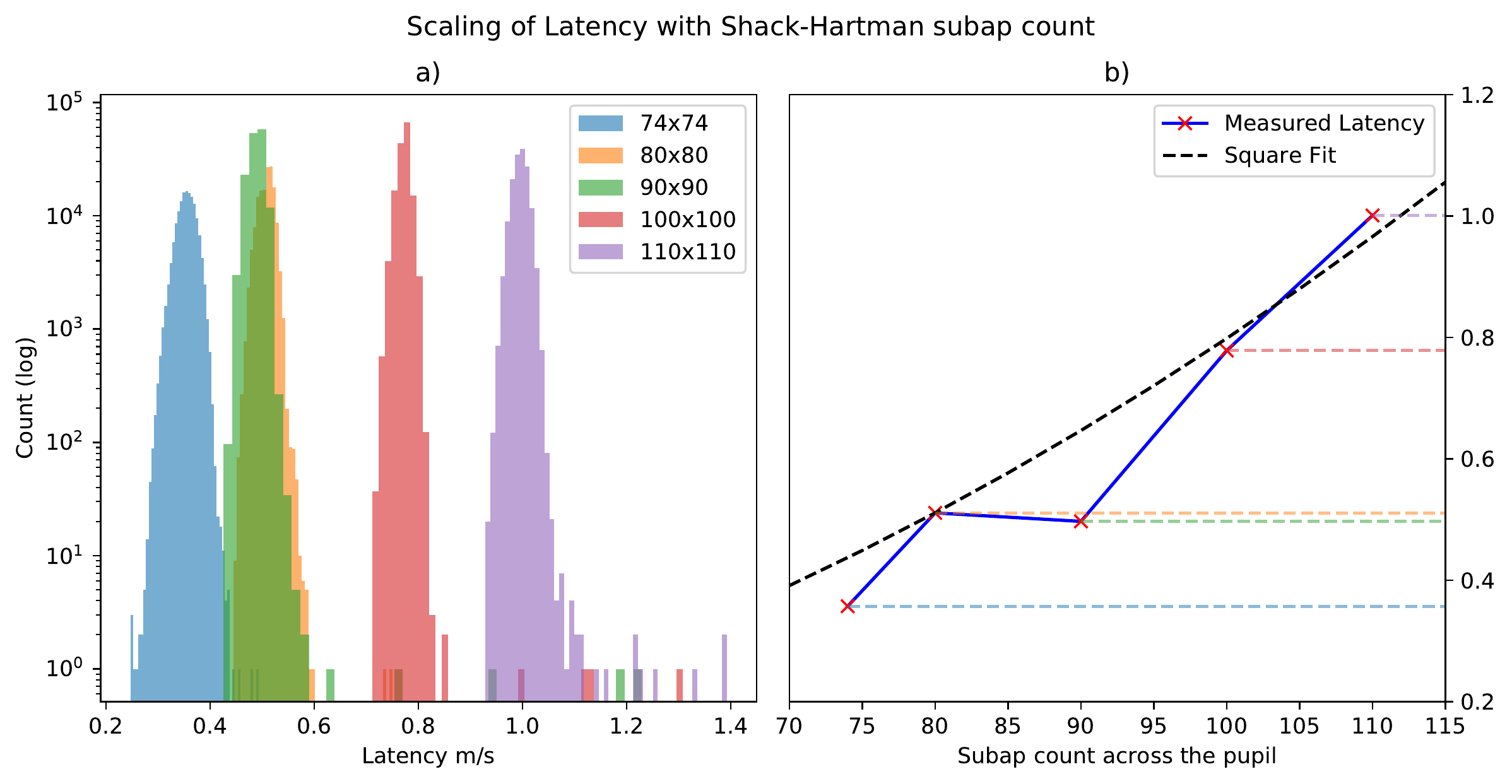}
    \caption{The scaling of latency with subapertures across the pupil for the Shack-Hartman type WFS. The latency is that for the entire RTC operation from pixels to DM commands as shown in Figure~\ref{fig:ao_diag}, measured from the time the last pixel arrives until the DM command is ready. The relevant values for the data are shown in Table~\ref{tab:ao_results}. }
    \label{fig:sh_scaling}
\end{figure}

\begin{figure}
    \includegraphics[width=1\columnwidth]{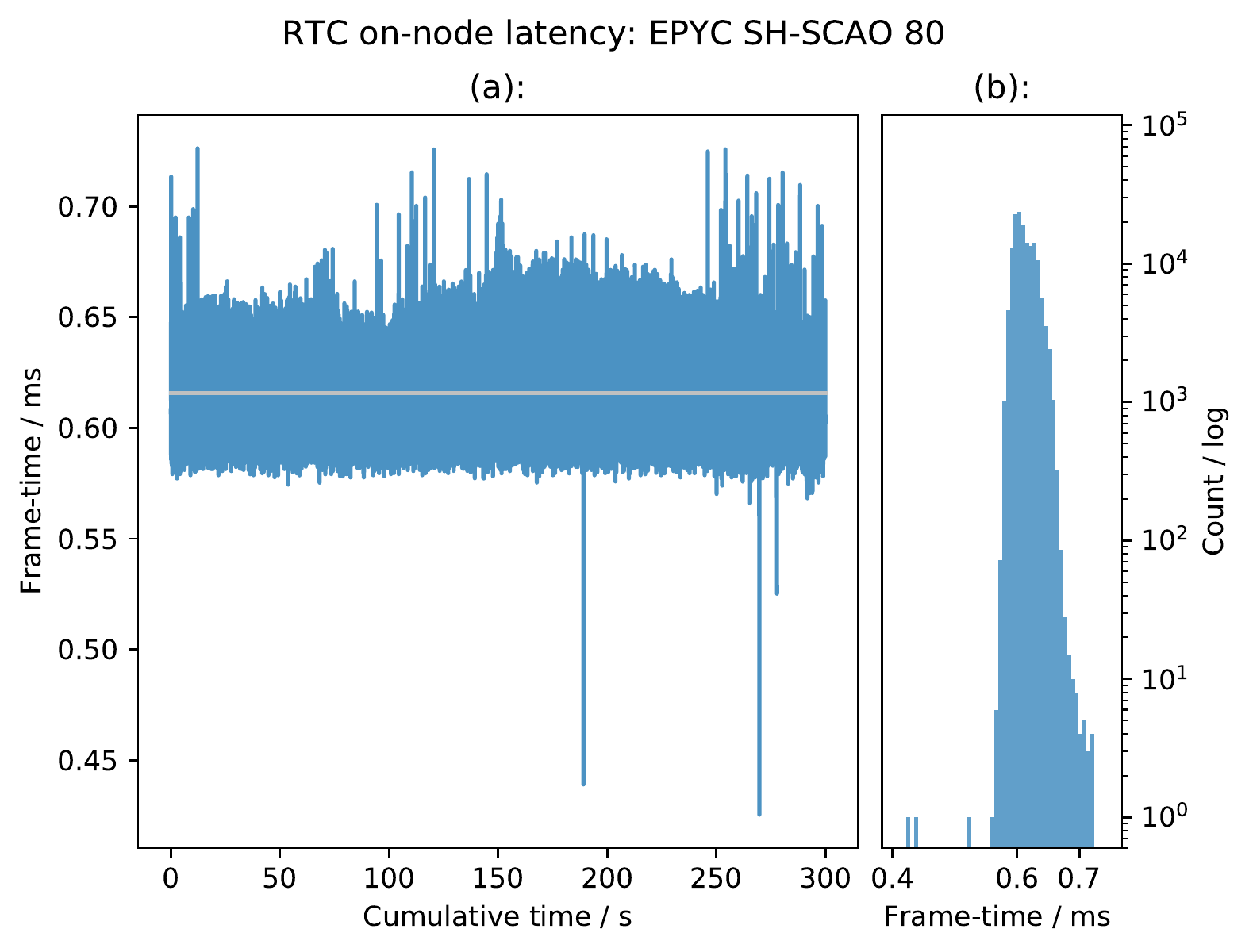}
    \caption{Frametime results for an SCAO setup using a Shack-hartman type WFS slope calculation with $80$ subapertures across the pupil for an AMD EPYC system as described in Section~\ref{sec:other_cpus}. Results shown are for $1.5\times10^5$ iterations at $500Hz$ for a total time of $300s$. The mean latency is $616 \pm 17 \mu s$. }
    \label{fig:epyc_scao}
\end{figure}

Figure \ref{fig:shwfs_scao} shows frame time and latency results for DARC running an SCAO RTC on an Intel Xeon Phi 7250 with an attached simulated camera using the UDP pixel streaming method as described in Section~\ref{sec:devel_RTC}. These results are for a SCAO setup with a single $80\times80$ SH-WFS with 4616 valid subapertures and an ELT-like M4 + M5 DM configuration with a total of 5318 actuators. The reconstruction therefore, is a single MVM of dimensions $5318\times9232$. There is 300 seconds worth of data corresponding to $1.5\times10^5$ frames at a framerate of 500Hz, the average latency is measured at $511 \pm 15 \mu s$. This compares favourably with similar results of $640 \pm 20 \mu s$ from \cite{Jenkins2018}, with a reconstructor of dimensions $5170\times9416$. The problem size used here has been adjusted to better reflect the potential dimensions of actual ELT instruments using more up to date information \citep{Biasi2016,Correia2018}.

Figure \ref{fig:pyrwfs_scao} shows frame time results for a similar AO setup as above but using a Pyr-WFS instead of a SH-WFS, with the differences between the WFS processing as described in Section~\ref{sec:devel_RTC}. There are 300 seconds worth of data corresponding to $1.5\times10^5$ frames at a framerate of 500Hz, and the average latency here is measured at $998 \pm 10 \mu s$. The dimensions of $100\times100$ pixels for the Pyr-WFS and $80\times80$ subapertures for the SH-WFS were chosen as these are the likely dimensions that would be used for real WFS on ELT-scale instruments. The SH-WFS dimensions result from technological limitations of the sensors being developed by ESO and the required dimensions of each subaperture \citep{Schreiber2018}. The Pyr-WFS dimensions were chosen based on what is being targeted for the ELT first light instrument HARMONI \citep{HARMONI2014} SCAO mode \citep{Schwartz2018}.

Figures~\ref{fig:sh_scaling} and~\ref{fig:pyr_scaling} show the latency scaling of both the SH-WFS RTC and Pyr-WFS RTC against sub-apertures/pixels across the pupil. Due to the use of a simulated camera as described in Section~\ref{sec:devel_RTC}, there are some unavoidable latency spikes that result from delays of the pixel transmission from the simulated camera. The mean latencies, RMS jitters and largest outliers are shown in Table~\ref{tab:ao_results} for data sets containing $1.5\times10^5$ samples at 500Hz. Table~\ref{tab:ao_results} also shows the RMS jitter and largest outliers for each case for a reduced subset of the data containing at least $2\times10^4$ continuous iterations. These reduced sets were chosen to eliminate any major outliers resulting from camera delays to give a better idea of the ``steady'' latency distribution.

For all of these results the input image sizes are kept constant for each type of WFS processing used, images of $800\times800$ pixels are used for the SH-WFS and 240x240 for the Pyr-WFS and either the number of pixels per sub-aperture is reduced to provide more sub-apertures or a smaller area within the input image is used for reduced numbers of sub-apertures. In this way the pixels received are kept constant between the different WFS and only the calibration, centroiding and reconstruction are affected by the different dimensions. We can see that for the Pyr-WFS the scaling matches very closely to a square fit, which is as expected, this is because the change in degrees of freedom in the reconstruction scales to the second power of the number of pixels across the pupil.

For the SH-WFS results shown in Figures~\ref{fig:sh_scaling} there is no clear fitting to a square fit and the case of $90\times90$ subapertures actually has a lower latency than the $80\times80$ subapertures case. We believe this is due to the reduced number of pixels per subaperture when using $90\times90$ subapertures. Because all of the SH-WFS tests use the same simulated camera image dimensions of $800\times800$ pixels, the $90\times90$ subapertures case is only using $8\times8$ pixels per subaperture compared to $10\times10$ pixels per subaperture for the $80\times80$ case. This reduces the perceived latency in two ways. Firstly by the reduced pixel processing required and secondly by the fact that the latency is measured as the time from when the last pixel arrives and due to the way this is measured, the timestamp is taken when the full $800\times800$ image has arrived which is later than when the processing completes $90\times90$ subapertures case.

If the latencies for the Pyr-WFS in Table~\ref{tab:ao_results} are compared directly with the latencies from the SH-WFS for the same WFS dimensions, we see that the SH-WFS overall results in reduced latencies. This is a result of the reduced pipe-lining efficiency of the Pyr-WFS vs. the SH-WFS as described in Section~\ref{sec:devel_RTC} and shown in Figure~\ref{fig:wfs_slopes}.

\subsection{Other many-core CPU systems} \label{sec:other_cpus}

Most of the results presented in this report are obtained from Intel Xeon Phi CPU systems as described in \citet{Jenkins2018}. However the Xeon Phi platform has been discontinued and so it is unlikely to be considered as a candidate for real AO RTC hardware. One of the main reasons for choosing a CPU based RTC is that the software and the optimisations made for many-core operation are not specific to a single CPU architecture or vendor. Figure \ref{fig:epyc_scao} shows frametime and latency results for DARC running an SCAO RTC on an AMD EPYC system with an attached simulated camera using the UDP pixel streaming method as described in Section~\ref{sec:devel_RTC}. This can be compared directly to the Xeon Phi results shown in Figure~\ref{fig:shwfs_scao}. There is 300 seconds worth of data corresponding to $1.5\times10^5$ frames at a framerate of 500Hz, the average latency is measured at $616 \pm 17 \mu s$.

The source code and RTC parameters are identical for the two different CPUs with the only differences coming from the compiler options and the configuration of the threading and NUMA aware memory allocation for the two different platforms. The EPYC system used for these results is an AMD EPYC 7351 dual socket system with 16 cores and 64GB of DDR4 2667 MHz memory per socket. The core topology of the EPYC CPUs is such that each CPU has four NUMA regions with each region having 4 CPU cores and 16GB of memory each. The RTC software uses the NUMA information of the CPU to allocate memory for the RTC control matrix on the nodes relevant to each CPU core. The Kernel and OS is tuned in a similar way to the Xeon Phi with the major differences being the OS, Ubuntu 16.04 for EPYC vs. CentOS for the Xeon Phi, and simultaneous multi-threading (Hyper-Threading) is turned on for the EPYC system as it provides better performance and allows 8 threads per NUMA node.

The maximum memory bandwidth of the EPYC system is $350GBs^{-1}$ by having 8 channels of DDR4 memory per socket and utilising NUMA aware software. This is less than the measured memory bandwidth of $480GBs^{-1}$ of the Xeon Phi 7250. It is therefore expected that the performance of the EPYC will be less than that of the Xeon Phi for the memory bandwidth bound RTC operations. However these results show that the software can be readily used on different CPU platforms and that performance is as expected based on the knowledge that the main RTC operations are memory bandwidth bound.

Other multi-socket CPU systems would also be suitable for ELT-scale AO RTC such as the Intel Xeon Scalable processors which in a quad socket configuration can provide higher maximum memory bandwidth than the Xeon Phi when the NUMA regions are taken into account. The multi-socket systems also benefit from generally running at a higher base CPU frequency than the Xeon Phi and so their single threaded performance is better. A dual-socket EPYC system with the required memory bandwidth can be purchased for a similar price to the Xeon Phi, making it the most likely substitute. For the Intel Xeon Scalable processors, due to their reduced memory channels per socket, a quad socket system would be required to match the memory bandwidth and this can increase the per node costs to over $4\times$ that of comparable EPYC or Xeon Phi systems.

Next generation AMD EPYC processors will introduce a new architecture \citep{AMDpresent} that simplifies the core topology of the system and will be built on a smaller 7nm process node to provide better energy efficiency. This involves introducing a 9-die architecture which includes 8 compute chiplets and a single I/O interface die such that each CPU core can access all memory channels equally. This is different to the current design where each of the 4 NUMA regions has 8 cores and 2 memory channels each and so only those 8 cores can access the full bandwidth of those 2 channels. This should reduce the relative complexity of NUMA memory management and allow more efficient interleaving of memory over all 8 memory channels which will likely reduce the latency of the EPYC results shown in Figure~\ref{fig:epyc_scao}. The memory for the next-generation processors is also likely to be clocked faster, at up to 3200MHz compared to the current maximum of 2667MHz, increasing memory bandwidth.

\section{Prototyping an MCAO and LTAO RTC} \label{sec:mcao_ltao_rtc}

\begin{table}
    \centering
    \caption{A comparison of the specifications of ELT-scale SCAO, MCAO and LTAO, the values are the most current known specifications for the HARMONI SCAO mode, the MAORY MCAO mode and the HARMONI LTAO mode respectively.}
    \label{tab:ao_compare}
    \begin{tabular}{llll} % four columns, alignment for each
        Mode & SCAO & MCAO & LTAO \\
    \hline
        Target Frame rate (Hz) & 1000 & 500 & 500\\
    \hline
        LGS number & 0 & 6 & 6\\
        LGS sub-aperture geometry & N/A & $80\times 80$& $80\times 80$\\
        LGS pixel geometry & N/A & $10 \times 10$&$10 \times 10$\\
        LGS total sub-apertures &N/A&$4616\times 6$&$4616 \times 6$ \\
        LGS image format &N/A&$800\times 800$& $800\times 800$\\
    \hline
        NGS number & 1 & 3 & 1\\
        NGS type & Pyramid& SH-WFS&SH-WFS\\
        NGS sub-aperture geometry & $100\times 100$ & $2 \times 2$ & $2 \times 2$ \\
        NGS pixel geometry & N/A & $100\times 100$&$100\times 100$\\
        NGS total sub-apertures &4616& $4\times 3$& $4$ \\
        NGS image format &$240\times 240$&$240\times 240$&$240\times 240$ \\
    \hline
        DM number &2&3&1\\
        Total DM modes &$5316+2$&$5316+2$&$5316+2$\\
         &$+176$&$+2\times 500$&$+6\times 2$\\
    \end{tabular}
\end{table}

As mentioned in Section~\ref{sec:mcao_ltao} MCAO and LTAO generally differ from SCAO by the number of WFSs and DMs used. Therefore in the context of the RTC they are both much more computationally demanding than the simple SCAO case. When prototyping an RTC architecture for MCAO and LTAO we decided to design it based on the ELT first-light instruments, the Multi-conjugate Adaptive Optics RelaY \cite[MAORY,][]{maory2018} and the HARMONI LTAO mode \citep{HARMONI2016}. For both of these instruments, the most demanding aspects will be the reconstruction of 6 laser guide star (LGS) WFSs operating at 500Hz, the parameters for these instruments and a comparison with an SCAO system are shown in Table~\ref{tab:ao_compare}. Targeting proposed ELT MCAO and LTAO instruments allows us to demonstrate more realistic test cases and puts better constraints on the design of the architecture.

One of the main benefits of designing a CPU based RTC lies in the flexibility and generality that the CPU architecture provides. The product of this is that the software optimisations and modifications detailed in \citet{Jenkins2018} for SCAO can be readily applied to different AO types such as MCAO and LTAO. Therefore the architecture we propose for these AO regimes is an extension of the SCAO case by scaling the software and hardware to match the increased computational complexity.

For the MAORY MCAO and HARMONI LTAO parameters detailed in Table~\ref{tab:ao_compare} the computational demands for each LGS WFS are similar to those of the SH-WFS SCAO case tested in Section~\ref{sec:scaoresults} with slightly more DM actuator DoF but targeting a reduced framerate. We therefore decided that for each LGS WFS the processing of the WFS images to reconstructed wavefronts can be done in a very similar way to the SCAO procedure in Section~\ref{sec:scaoresults} by processing a single LGS WFS per processing node.

The NGS parameters demand far fewer computational resources than the SCAO case demonstrated on a single node in Section~\ref{sec:scaoresults}, therefore we propose that the NGS WFS processing, 3 NGS for MAORY and 1 for HARMONI, can be achieved by a single instance of DARC on a single processing node. The partial DM commands resulting from these calculations then need to be summed together to produce a single DM command vector for all of the required DMs. This will be achieved by having a separate ``master'' processing node which can receive partial DM commands from the reconstruction nodes and perform any post-processing that may be required before delivering the final actuator commands to the DMs.

\begin{figure}
    \includegraphics[width=1\columnwidth]{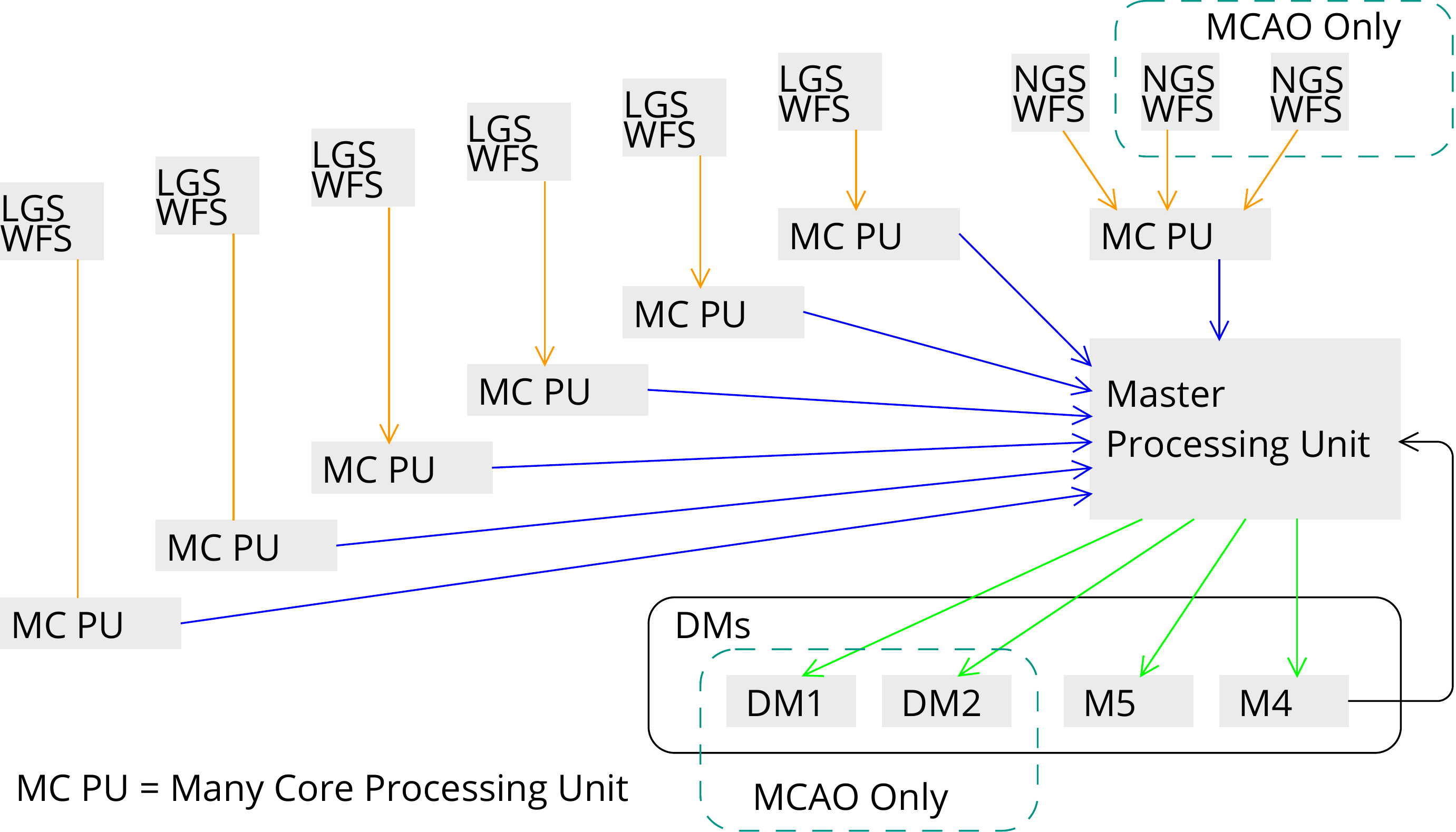}
    \caption{The proposed architecture for the MCAO and LTAO multi node RTC. The 6 LGS WFSs are each processed by a single many-core node. The NGS are all processed on the same node, three for MAORY and one for HARMONI. The master node sends out the final DM commands once it has finished summing and processing the partial vectors. The master node should also able to receieve feedback from the ELT M4 to integrate the actual M4 shape used in the next command. }
    \label{fig:mcaoltao_arch}
\end{figure}

\begin{figure}
    \includegraphics[width=1\columnwidth]{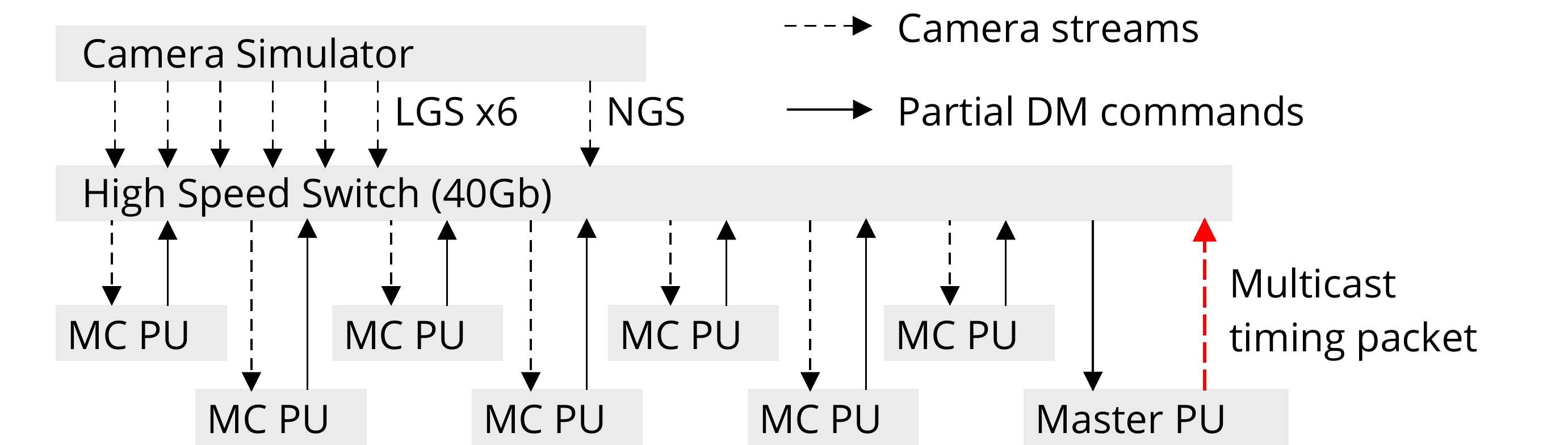}
    \caption{The lab test setup for the MCAO and LTAO architecture. The simulated camera streams 6 LGS WFSs and either 1 or 3 NGS WFSs through a high speed switch to the processing units. The many-core processing units send their partial DM commands calculated from the individual WFSs to the master node for summing and further processing. For the tests to determine the overall latency of the system, the master node multicasts a timing packet for all other nodes to time stamp the end of frame. }
    \label{fig:mcaoltao_test}
\end{figure}

Figure~\ref{fig:mcaoltao_arch} shows the prototype architecture for MCAO/LTAO RTC using the ideas described above. There are a total of 7 reconstruction nodes to process all the WFSs required for the either the MAORY MCAO case or the HARMONI LTAO case. An 8th master node is used for summing the partial results from each reconstruction node and processing them for sending to the DMs. Due to the way the ELT M4 will operate \citep{Xompero2018}, the correction applied by M4 will not necessarily be the same as that which is delivered to M4 from the RTC. It will therefore be necessary for the RTC of an ELT instrument to receive feedback from M4 such that it can incorporate the actual correction applied for the previous iteration. In our prototype architecture, this will be processed by the master processing unit, shown in Figure~\ref{fig:mcaoltao_arch}.

Figure~\ref{fig:mcaoltao_test} shows a lab test set up derived from the prototype architecture shown in Figure~\ref{fig:mcaoltao_arch}. It shows the simulated camera delivering the required camera streams to each reconstruction node over a high speed network. The master processing node communicates with the reconstruction nodes over a separate high speed network on the same physical switch but using a different VLAN and different network interconnects to the camera streams. For our lab test setup the full RTC latency is calculated on each reconstruction node by measuring the time between when it receives the last pixel from its camera stream to the time it receives a timing packet from the master which is multicast to the reconstruction nodes when it has completed the current frame. As the cameras use the proposed ESO MUDPI packet they stamp each image with a frame number which is then propagated through to the master and back through the timing packet to be able to match the correct DM command to the camera frames.

% \subsection{UDP cameras simulator setup for MCAO/LTAO}

% \textit{(Detail about how the simulator sends the 7 different camera streams)}

\subsection{Results of testing the prototype}

\begin{figure}
    \includegraphics[width=1\columnwidth]{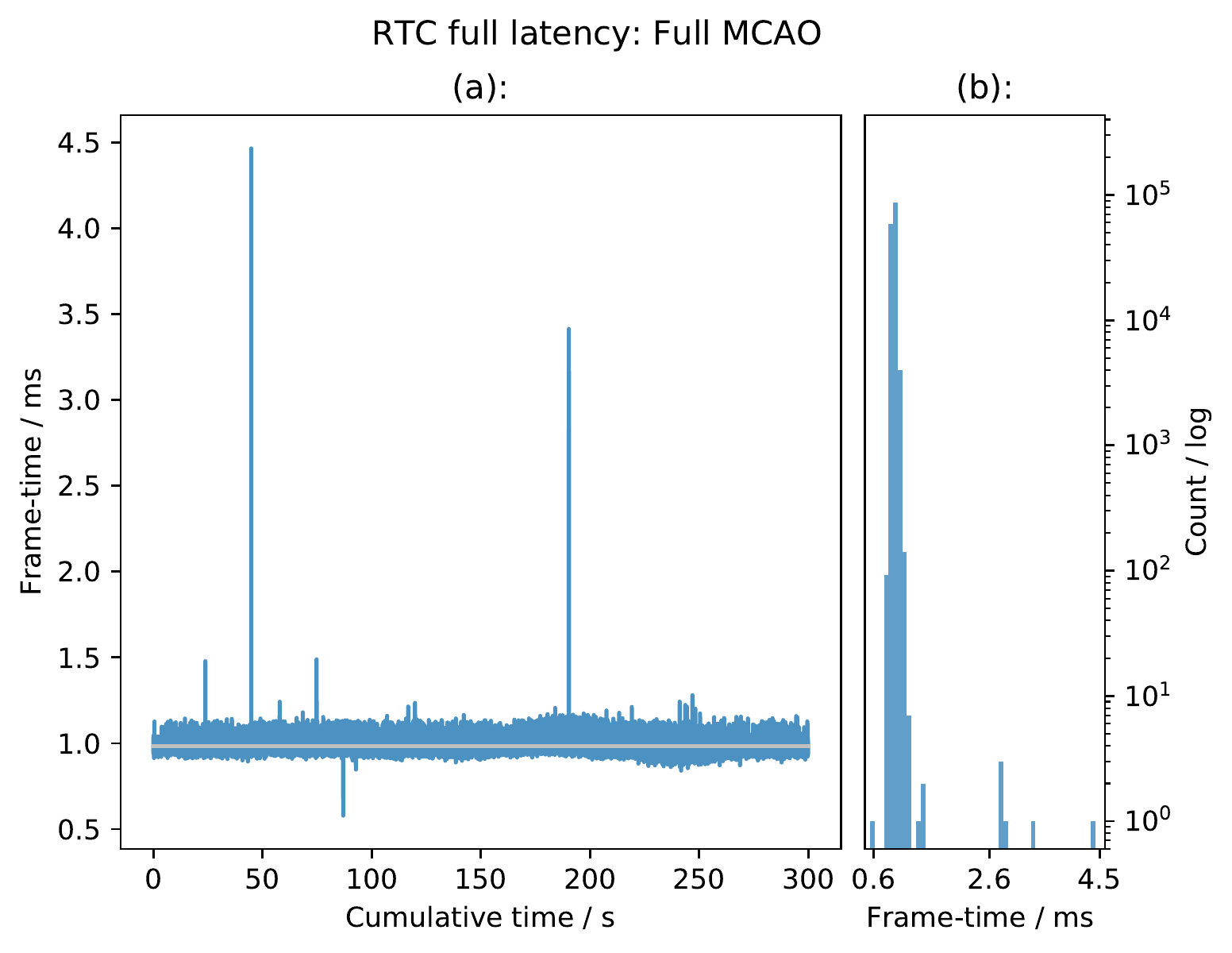}
    \caption{Latency results for the full MCAO setup as described in Section~\ref{sec:mcao_ltao} and Table~\ref{tab:ao_compare}. This is for $1.5\times10^5$ iterations at $500Hz$ corresponding to a total cumulative time of $300s$. This is a measure of the time between when the last pixel arrives at a reconstruction node and when it receives the timing packet from the master node. The large outliers are result of delays from the CPU based simulated cameras, which is compounded by the fact that this timing data includes delays from all 7 simulated camera streams. }
    \label{fig:full_mcao}
\end{figure}

\begin{figure}
    \includegraphics[width=1\columnwidth]{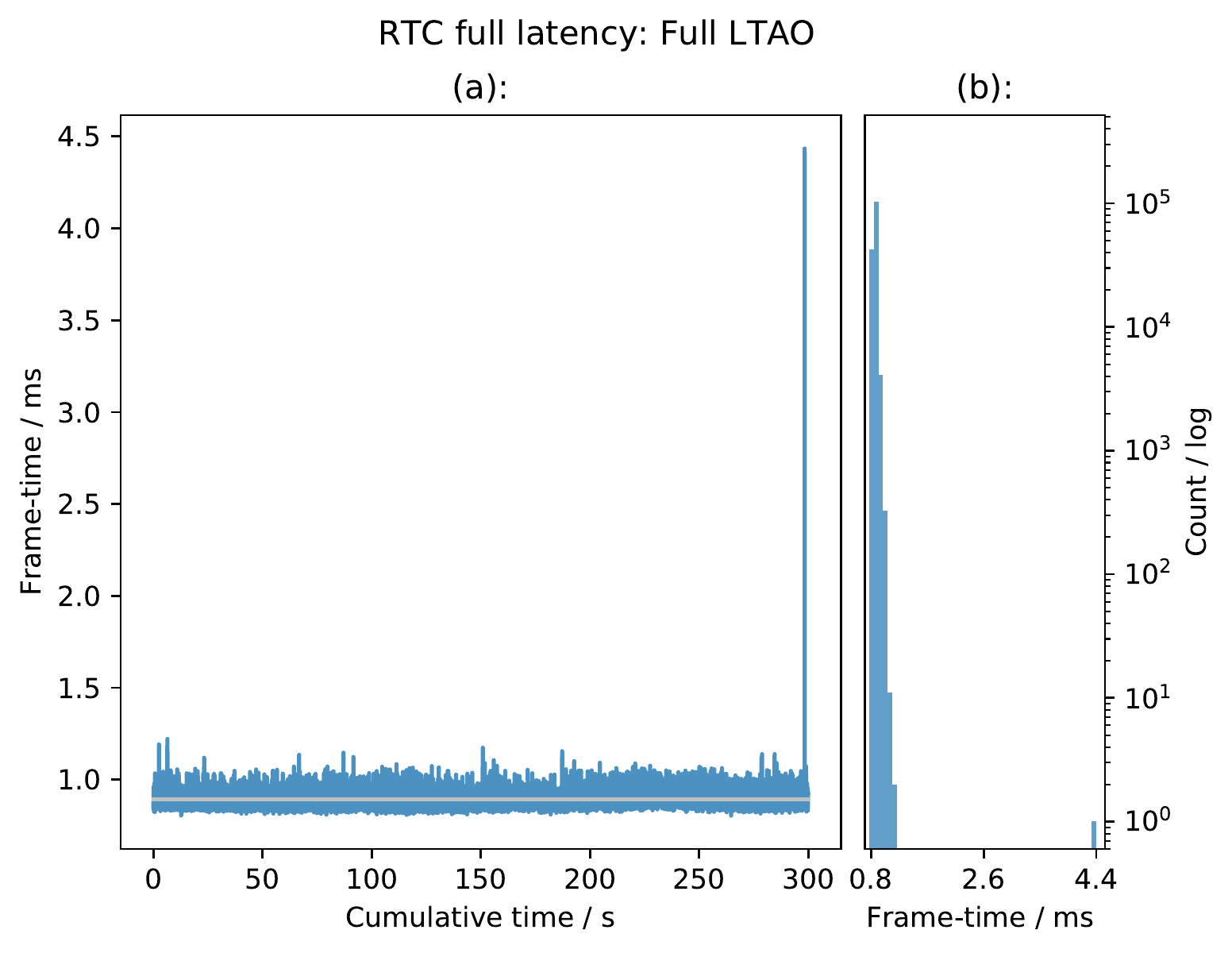}
    \caption{Latency results for the full LTAO setup as described in Section~\ref{sec:mcao_ltao} and Table~\ref{tab:ao_compare}. This is for $1.5\times10^5$ iterations at $500Hz$ corresponding to a total cumulative time of $300s$. This is a measure of the time between when the last pixel arrives at a reconstruction node and when it receives the timing packet from the master node. The large outliers are result of delays from the CPU based simulated cameras, which is compounded by the fact that this timing data includes delays from all 7 simulated camera streams. }
    \label{fig:full_ltao}
\end{figure}

\begin{table}
    \centering
    \caption{Latency, RMS jitter and largest outliers results for all of the data presented in this report. For all results other than LTAO with buffer swap the a) columns correspond to results from $1.5\times10^5$ continuous iterations at $500Hz$ for a total time of $300s$ for each test case. The b) columns correspond to results from a subset of no less than $2\times10^4$  continuous iterations chosen from the larger a) data sets for a toal time of $40s$ each. The b) column subsets were chosen to avoid any large outliers that result from simulated camera delays to give a better representation of the ``steady'' latency. The LTAO with buffer swap results are for $1.5\times10^4$ continuous iterations at $500Hz$ for a total time of $30s$, there are no ``steady'' results for this case as the outliers here are a result of the parameter swap itself and not from an external factor. }
    \label{tab:ao_results}
    \begin{tabular}{lrrrrr}
        & Mean & \multicolumn{2}{r}{RMS} & \multicolumn{2}{r}{Largest} \\
         & Latency  & \multicolumn{2}{r}{ Jitter $(\mu s)$} & \multicolumn{2}{r}{Outlier $(\mu s)$} \\
        AO Mode&$(\mu s)$&a)&b)&a)&b)\\
    \hline
        Pyr-WFS $80\times80$   &  609 & 11 & 11 & 898  & 653  \\
        Pyr-WFS $90\times90$   &  796 & 16 & 16 & 1051 & 862  \\
        Pyr-WFS $100\times100$ &  998 & 10 &  9 & 1667 & 1049 \\
        Pyr-WFS $110\times110$ & 1209 & 12 & 12 & 1433 & 1292 \\
        Pyr-WFS $120\times120$ & 1450 &  8 &  8 & 1886 & 1508 \\
    \hline
        SH-WFS $74\times74$    &  357 & 18 & 16 & 495  & 436  \\
        SH-WFS $80\times80$    &  511 & 15 & 15 & 754  & 589  \\
        SH-WFS $90\times90$    &  497 & 14 & 14 & 1237 & 559  \\
        SH-WFS $100\times100$  &  778 & 12 & 11 & 1314 & 821  \\
        SH-WFS $110\times110$  & 1001 & 14 & 14 & 1397 & 1065 \\
    \hline
        Full MCAO              &  985 & 33 & 29 & 4465 & 1235 \\
        MCAO telemetry         & 1085 & 32 & 30 & 2988 & 1466 \\
        MCAO POLC              & 1090 & 45 & 44 & 2880 & 1312 \\
        MCAO 6 LGS             &  992 & 47 & 46 & 3498 & 1143 \\
        MCAO 5 LGS             &  979 & 46 & 44 & 2870 & 1261 \\
        MCAO 4 LGS             &  969 & 45 & 45 & 3305 & 1285 \\
        MCAO 3 LGS             &  943 & 43 & 42 & 2817 & 1182 \\
        MCAO 2 LGS             &  951 & 42 & 43 & 2858 & 1119 \\
    \hline
        Full LTAO              &  894 & 29 & 28 & 4434 & 1174 \\
    \hline\hline
        LTAO with buffer swap        & 1034 & 31 & -- & 1949 & --   \\
    \end{tabular}
\end{table}

When testing the MCAO prototype described in Section~\ref{sec:mcao_ltao_rtc} certain considerations needed to be made with regards to the timing of the individual frames. Ideally we would want to measure the time from last pixel into the RTC until the time the DM command is ready, however we discovered that for a multi-node CPU based architecture it is difficult to synchronise the clocks between nodes with enough precision using our available hardware such that timestamps generated on each node can be directly compared. Instead we have to rely on only comparing timestamps generated on individual nodes and so the full RTC latency is calculated as the time between a reconstruction node receiving the last pixel from its camera stream and the time it receives a timing packet from the master node indicating that the final DM command is ready, which does result in a slightly pessimistic measurement.

Ideally the full RTC latency would be measured externally with a device that is able to record the time between when the camera finishes sending a frame and the time when the corresponding DM command is received from the master node. This process would make the RTC latency measurement easier however the method described above is no less valid as a measure of the RTC latency.

For all the MCAO and LTAO tests described in this report, the system architecture used is shown in Figure~\ref{fig:mcaoltao_arch} and network interconnects are as shown in Figure~\ref{fig:mcaoltao_test}. Each of the 7 reconstruction nodes uses a single instance of the DARC software to receive an $800\times800$ pixel camera stream which it process from pixels to a partial DM command for that WFS. This processing is done in a very similar way to the SH-WFS SCAO case as descibed in \citet{Jenkins2018} and in Section~\ref{sec:scaoresults}, the main differences are the way in which the reconstruction matrix is constructed and that the DM software library is used to send the partial DM to the master node. The master node itself runs a separate instance of DARC which receives the partial DM commands, sums them together, processes the result and once finished it sends a timing packet to the other nodes.

Figure \ref{fig:full_mcao} and Figure~\ref{fig:full_ltao} shows RTC latency plots for the MCAO and LTAO test cases respectively. They show the timing data from one of the reconstruction nodes of the 7 during full operation, the timing data also includes the transmission and receiving time of the DM timing packet and can therefore be considered slightly pessimistic. The specification used for each are shown in Table~\ref{tab:ao_compare} and the mean latency of the MCAO case is $985 \pm 33 \mu s$ and $894 \pm 29 \mu s$ for the LTAO case. It can be seen that there are 2 major outliers in the latency for the MCAO test and one for the LTAO test which are a result of delays introduced from the CPU based simulated cameras. The number of frames losses however is acceptable considering that there is at worst one frame drop per $150s$ total integration time.

As well as testing the full 8 node MCAO RTC we also tested the architecture with different numbers of reconstruction nodes combined with the master node. Table~\ref{tab:ao_results} shows the results for the MCAO case where there are less than the full number of reconstruction nodes for the MAORY specification. Results are shown for the cases where there are 2, 3, 4, 5 and 6 LGS reconstruction nodes feeding partial DM commands to the master node. As can be seen from the results the total latency varies only slightly by the reduction in processing nodes, showing that the solution is scalable at least up until the desired number of nodes for ELT-scale MCAO.

\subsection{Effect of on-the-fly changes to RTC parameters on latency} \label{sec:param_swap}

\begin{figure}
    \includegraphics[width=1\columnwidth]{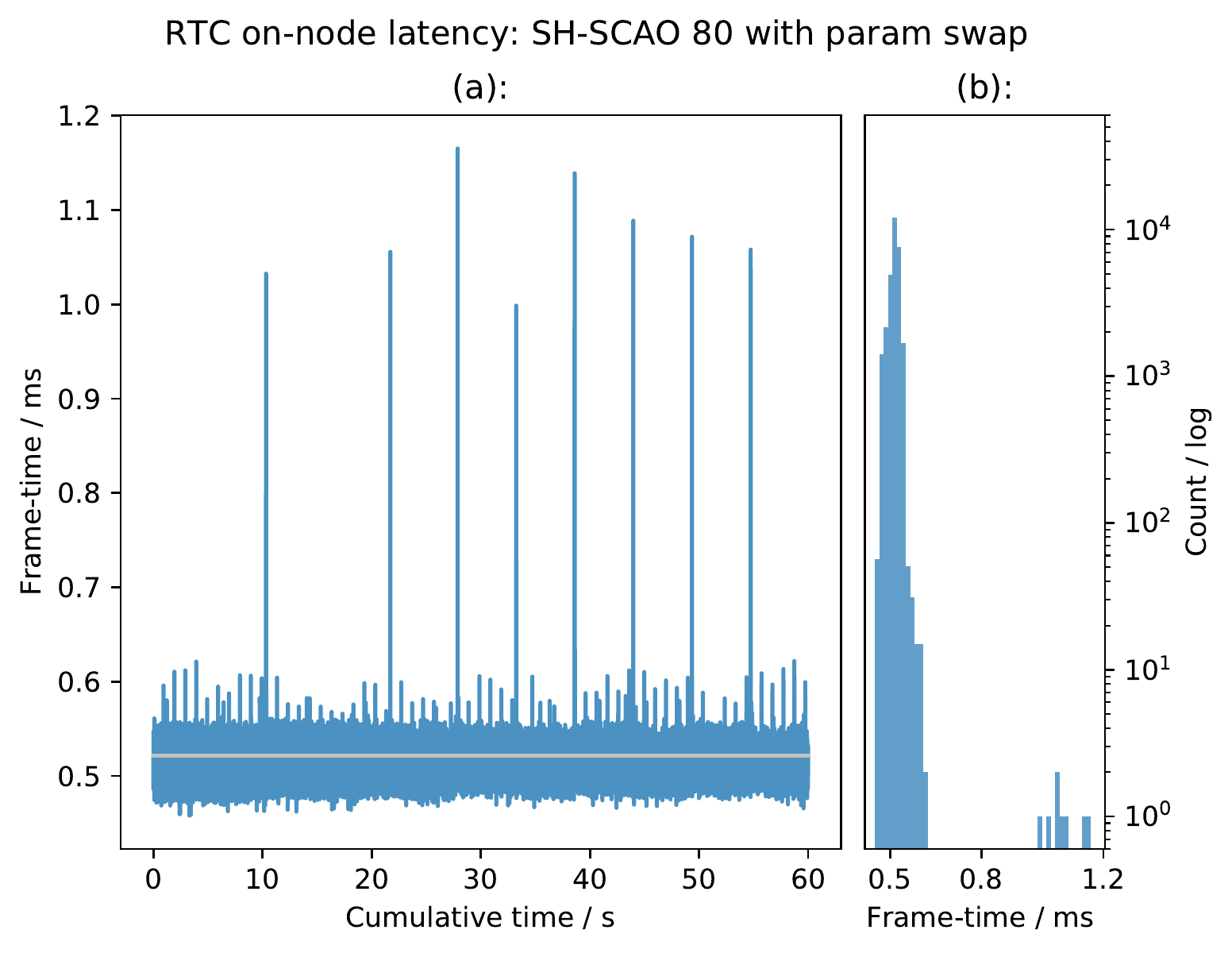}
    \caption{Latency results for an SCAO setup as described in Section~\ref{sec:scaoresults} whilst two types of buffer swap are performed as described in Section~\ref{sec:param_swap}. As can be seen, the buffer swap causes a $\approx650\mu s$ spike to the latency whenever it is performed. This is for $3\times10^4$ iterations at $500Hz$ corresponding to a total cumulative time of $60s$ to highlight the 10 second transfer time of the first two parameter changes of the control matrix. }
    \label{fig:scao_param}
\end{figure}

\begin{figure}
    \includegraphics[width=1\columnwidth]{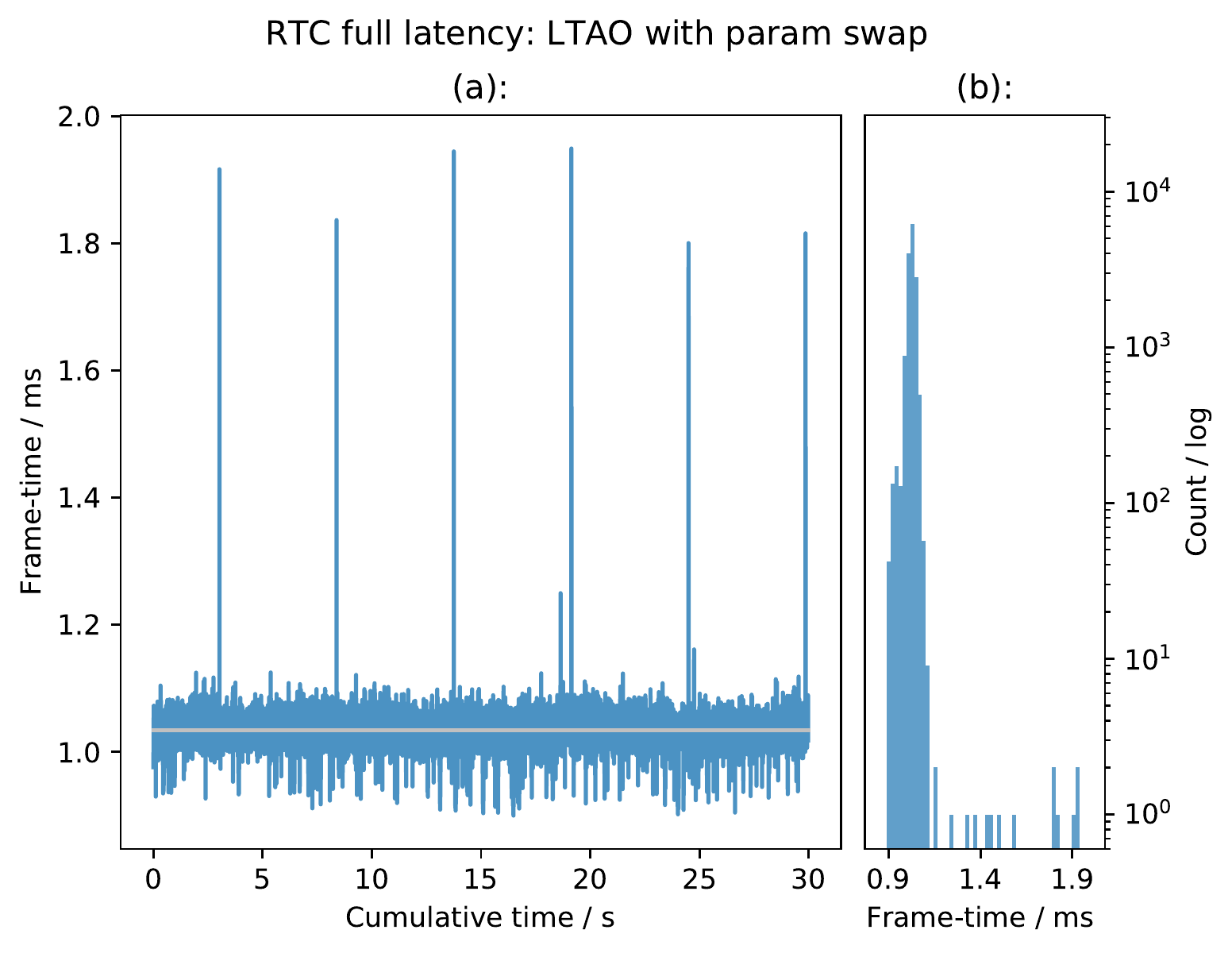}
    \caption{Latency results for an LTAO setup as described in Section~\ref{sec:mcao_ltao} whilst a periodic buffer swap is performed as described in Section~\ref{sec:param_swap}. As can be seen, the buffer swap causes a $\approx800\mu s$ spike to the latency whenever it is performed. This is for $1.5\times10^4$ iterations at $500Hz$ corresponding to a total cumulative time of $30s$ to highlight the disturbance every $5s$. }
    \label{fig:ltao_param}
\end{figure}

An important aspect of any real on-sky RTC is the ability to change parameters during operation, for example to update the matrix used in the reconstruction process or to update the reference centroids. DARC has the ability to set parameters through two different means, either through a command line utility called ``darcmagic'' which can be used to change simple parameters such as string and scalar values, or through a Python interface which can also change more complex parameters such as the arrays and matrices. DARC uses a double buffer approach to handle parameter switching, the buffers contain all the necessary parameters for RTC operation and whilst one of the buffers is read by the RTC during operation the second can be modified to include any required new values without affecting the running processes.

Once the necessary changes have been made to the second buffer, DARC performs a buffer swap which causes it to start reading values from the second buffer instead of the first. The process that DARC uses to handle a buffer swap involves a flag in the main processing loop which instructs the first thread that begins a new frame that a buffer swap is required and that thread performs some checks, updates some information and then replaces the buffer pointer. Because all of the processing threads will read from the buffer this needs to be thread safe and so all other threads are temporarily blocked while the buffer is swapped.

The double buffering of the parameters reduces the effects of a parameter change by allowing the majority of the change to happen during general operation without affecting latency. However on a platform like the Xeon Phi which excels on multi-threaded performance, the single threaded buffer swap can have a noticeable impact on the latency of the frame during the swap. Figure~\ref{fig:scao_param} shows the effect of a buffer swap on an SCAO system set up as described in Table~\ref{tab:ao_compare}. The first 2 latency spikes seen are due to changing the control matrix of size $5318\times9232$ which takes approximately 10 seconds for the transfer to happen. The other latency spikes are due to a periodic change of a single value parameter every five seconds. The size of the latency spikes shows that for different size parameters the amount of jitter introduced is the same and the spike only occurs once the internal RTC buffer is actually swapped.

Figure~\ref{fig:ltao_param} shows the latency for an LTAO type system set up as described in Table~\ref{tab:ao_compare} whilst a buffer swap is set to occur on one of the reconstruction nodes every five seconds. There is a clear impact on the latency which corresponds to a $\approx800\mu s$ spike to latency when the buffer swap occurs. Here the relative size of the latency spikes is increased from the SCAO case above as this is the full LTAO RTC system as described in Section~\ref{sec:mcao_ltao}. This essentially results in a frame drop for every buffer swap due to the average latency being $1034\mu s$ for this particular case.

We believe that the relatively large effect of a buffer swap on the latency of the LTAO system in this case is partly caused by the fact that the Xeon Phi has relatively poor single threaded performance and partly because of the need for all reconstruction nodes to be synchronised by the master node, compounding any adverse effects of the swap. The parameter change performed here was for a single valued scalar parameter, however due to the double buffered approach of DARC, changing more complex parameters such as arrays or matrices shouldn't have any more impact on the latency as all copying of data can occur concurrently with RTC operations.

\subsection{Effect of streaming RTC telemetry on latency} \label{sec:telemetry}

\begin{figure}
    \includegraphics[width=1\columnwidth]{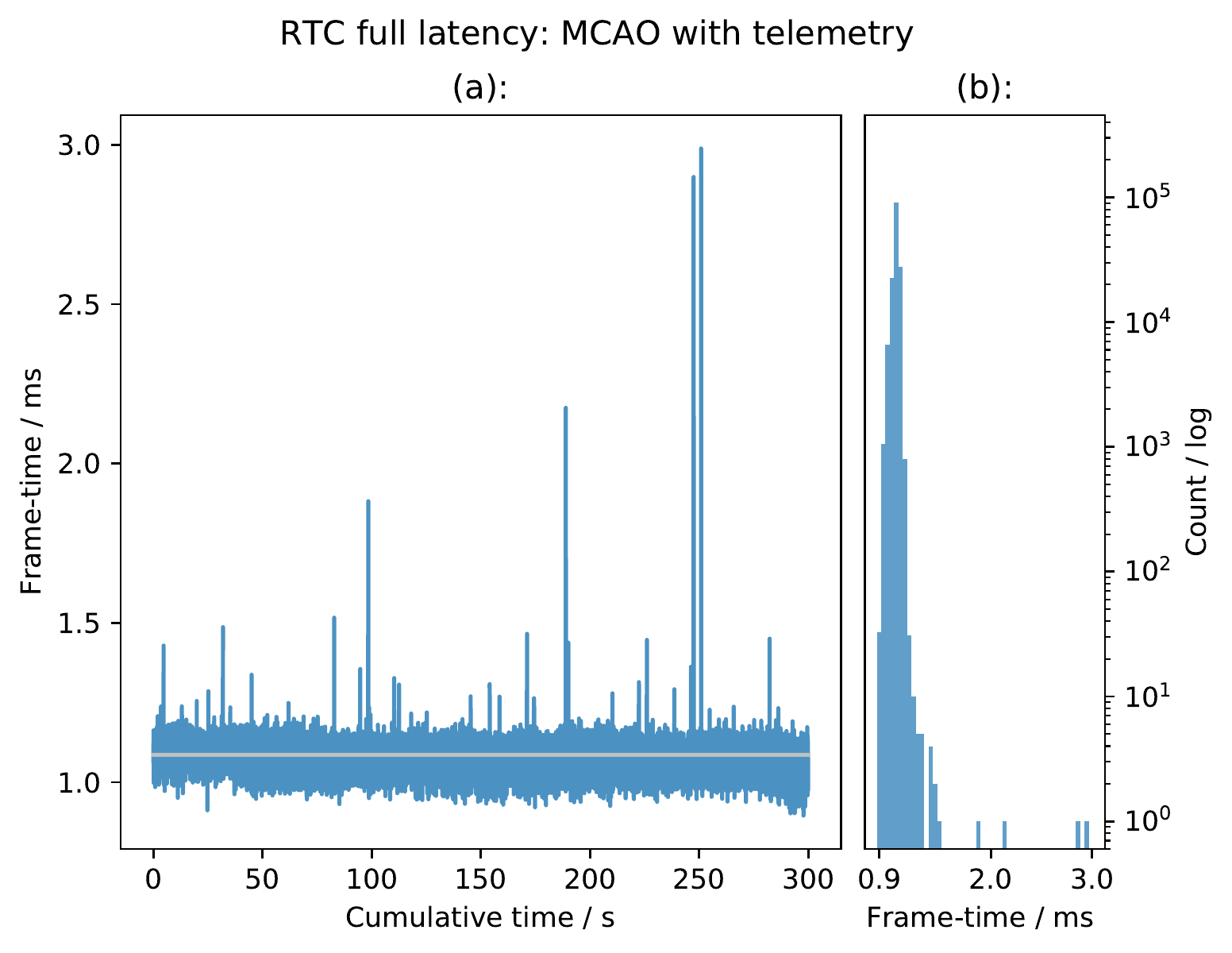}
    \caption{Latency results for an MCAO setup as described in Section~\ref{sec:mcao_ltao} whilst both centroid and DM command telemetry is taken for all nodes as described in Section~\ref{sec:telemetry}. This is for $1.5\times10^5$ iterations at $500Hz$ corresponding to a total cumulative time of $300s$. }
    \label{fig:mcao_telem}
\end{figure}

Another very important aspect of AO RTC operation involves the streaming of telemetry during operation, either for concurrent processing so as to update the reconstruction matrices or reference centroids or purely for saving data to disk for later analysis. DARC employs circular buffers to store telemetry when requested during operation and also has the capability to read from these buffers and send the telemetry to wherever is necessary. All the timing data used in this report is gathered by using the telemetry streaming functionality of DARC. There are 3 buffers which are read for every timing measurement used in this report; an RTC time buffer which stores frame times, an RTC status buffer which stores various status information and and RTC DM time buffer which stores the timing data for the receipt of the timing packet from the master node.

The status buffer is used to retrieve the timestamps for when the last pixel has arrived and also the timestamp for when each node has delivered its partial DM command. The DM time buffer is populated by a process which listens for packets from the master node and takes a time stamp on arrival. A common iteration number between the two buffers originating from the simulated camera is used to synchronise the data. In this way we can match up the DM command timing packet sent from the master to the image frame received from the simulated camera and calculate the full RTC latency.

Figure~\ref{fig:mcao_telem} shows the RTC latency for an MCAO setup for a case when slope telemetry and partial DM command telemetry is also streamed from the reconstruction nodes during operation. The mean latency is measured at $1085 \pm 32 \mu s$ and there are several outliers which result from delays due to the simulated camera streams. There are also a number of relatively small outliers in this data, $<1.5ms$, compared to the case without slope and DM telemetry which results from the taking of telemetry itself.

\subsection{Effect of pseudo-open loop control on latency} \label{sec:polc}

\begin{figure}
    \includegraphics[width=1\columnwidth]{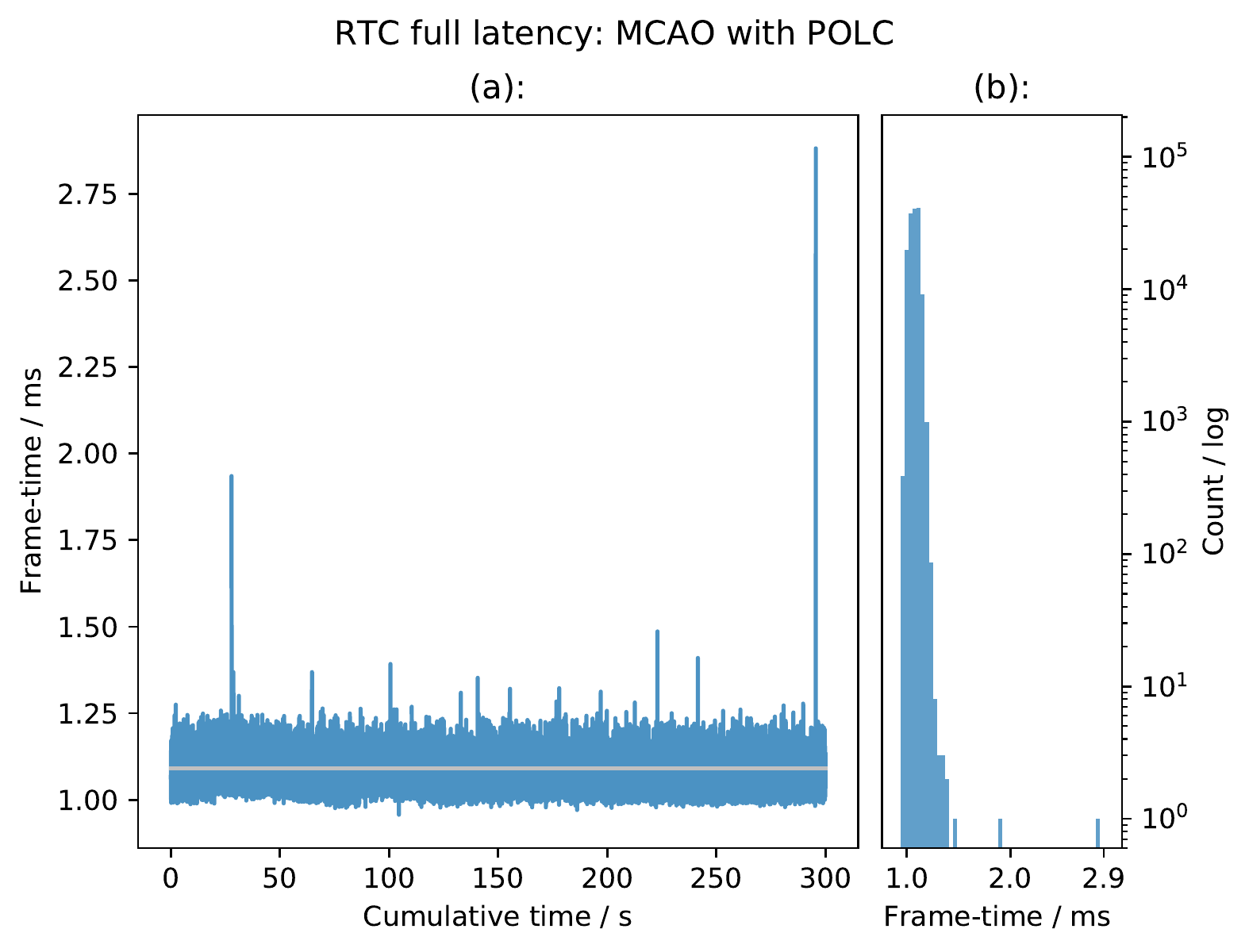}
    \caption{Latency results for an MCAO setup as described in Section~\ref{sec:mcao_ltao} whilst implicit POLC is computed on the master processing node as described in Section~\ref{sec:polc}. This is for $1.5\times10^5$ iterations at $500Hz$ corresponding to a total cumulative time of $300s$. }
    \label{fig:mcao_polc}
\end{figure}

All of the types of AO used in this report would generally be used in closed-loop operation, that is, the atmospheric wavefronts are corrected before the residual phase error is measured by the WFSs. This approach means that the wavefront phase errors measured by the WFSs are smaller than those measured in open loop and so the WFSs can be tuned for finer precision. Also, during closed-loop operation, errors in the measurement, correction, and reconstruction can be dynamically removed by the feedback of the system. The downside to closed-loop AO is that the slopes measured by the WFS no longer give a measurement of the actual atmospheric wavefront phase. The slopes measured in open-loop can be used to retrieve information about the atmospheric conditions which are required for reconstruction algorithms that take the current atmospheric statistics into account.

To get around the lack of open-loop slopes in closed-loop operation, it is possible to reconstruct pseudo-open loop (POL) \citep{Piatrou2005} slopes from the DM commands and resulting closed-loop slopes. For reconstruction algorithms that rely on POL control (POLC) there are 2 ways in which the POL slopes can be incorporated into the final reconstruction result. They can either be calculated explicitly and used directly in the algorithms or the effects of POLC can be incorporated into the final reconstruction implicitly without first calculating the actual POL slopes. The benefits of implicit POLC are massively reduced computational requirements. Explicit POLC requires the POL slopes to be computed before the MVM to calculate DM commands can be computed, however implicit POLC only needs the final DM command as calculated by the master node and so the POLC computation can be done there.

We have implemented the implicit POLC calculation for the MCAO and LTAO operation of DARC on the master node. The POLC is calculated for the next frame after the DM command is ready and so it should have minimal impact on the overall latency. The summing of partial DM commands on the master node is calculated by a single thread and so there are enough computational resources remaining to calculate the implicit POL, which is a single MVM, without affecting latency. Figure~\ref{fig:mcao_polc} shows the RTC latency for an MCAO setup for a case when the master node is performing POLC computation using 32 threads. The mean latency is measured at $1090 \pm 45 \mu s$ and there is a single large outlier which results from delays due to the simulated camera streams.
% !TEX root = ../main.tex
\section{Conclusions}
We have presented an update on the work and results in \citet{Jenkins2018} and demonstrated a prototype many-core CPU RTC architecture for MCAO and LTAO which builds on the previously presented work. We have shown that latencies of $<1000\mu s$ can be achieved for both MCAO and LTAO using this architecture with CPU based simulated cameras. We have demonstrated that running the DARC RTC software on Xeon Phi processors can achieve ELT SCAO latencies of less $<600\mu s$ for Shack-hartman WFS processing and $<800\mu s$ for Pyramid WFS processing. We have also demonstrated latencies of less $<700\mu s$ for Shack-hartman WFS processing on an AMD EPYC system, showing the generality of the software and techniques. The effects of parameter switching, telemetry streaming and implicit POLC computation have also been shown for the MCAO and LTAO cases with minimal effects on the overall latency. The CPU based simulated cameras have been shown to introduce large outliers in the latency distributions which are an unavoidable consequence of the hardware used.

As shown in \citet{Jenkins2018} the Xeon Phi processor is an example of a many-core CPU system with high-bandwidth memory and so the software and system architectures described in this report are readily transferable to other many-core CPU systems with the required computational and memory bandwidth specifications. More general many-core CPU systems could also achieve better latency and jitter due to the relatively weak single threaded performance of the Xeon Phi which greatly effects serial processes such as receiving pipe-lined pixel streams over network interfaces.

\section*{Acknowledgements}

Real-time AO work by the Durham group is supported by the EU H2020 funded GreenFlash project, ID 671662, under FETHPC-1-2014, the UK Science and Technology Facilities Council consolidated grant ST/P000541/1, and an STFC PhD studentship, award reference 1628730.

%%%%%%%%%%%%%%%%%%%%%%%%%%%%%%%%%%%%%%%%%%%%%%%%%%



\bibliographystyle{mnras}
\bibliography{refs} % if your bibtex file is called example.bib

%%%%%%%%%%%%%%%%%%%%%%%%%%%%%%%%%%%%%%%%%%%%%%%%%%

%%%%%%%%%%%%%%%%% APPENDICES %%%%%%%%%%%%%%%%%%%%%

% \appendix

% \section{Some extra material}

% If you want to present additional material which would interrupt the flow of the main paper,
% it can be placed in an Appendix which appears after the list of references.

%%%%%%%%%%%%%%%%%%%%%%%%%%%%%%%%%%%%%%%%%%%%%%%%%%

% Don't change these lines
\bsp	% typesetting comment
\label{lastpage}
\end{document}